\documentclass[a4paper,11pt]{article}
\pdfoutput=1 

\usepackage{jheppub} 
\makeatletter
\DeclareRobustCommand*{\bfseries}{%
  \not@math@alphabet\bfseries\mathbf
  \fontseries\bfdefault\selectfont
  \boldmath
}
\makeatother

\usepackage{calc}
\usepackage{rotating}
\usepackage[english]{babel}
\usepackage{graphicx}
\usepackage{subfig}
\usepackage{float}
\usepackage{amsmath}
\usepackage{amssymb}
\usepackage{amsthm}
\usepackage{latexsym}
\usepackage{dcolumn}
\usepackage{hyperref}
\usepackage{slashed}
\usepackage{mciteplus}
\newcommand{\newc}{\newcommand*}

\long\def\begincomment#1\endcomment{%
        \begingroup\sf\baselineskip12pt#1\endgroup}

\newc{\etal}{\textrm{et al.}} 
\newc{\eg}{\textrm{e.g.}} 
\newc{\ie}{\textrm{i.e.}}
\newc{\etc}{\textrm{etc}}
\newc\vs{\textrm{vs.}}
\newc{\cl}{\rm {C.L.}}

\newc{\ev}{\ensuremath{\,\mathrm{eV}}}
\newc{\kev}{\ensuremath{\,\mathrm{keV}}}
\newc{\mev}{\ensuremath{\,\mathrm{MeV}}}
\newc{\gev}{\ensuremath{\,\mathrm{GeV}}}
\newc{\tev}{\ensuremath{\,\mathrm{TeV}}}
\newc{\MeV}{\mev} 
\newc{\TeV}{\tev}
\newc{\invpb}{\ensuremath{/\text{pb}}}
\newc{\invfb}{\ensuremath{\,\textrm{fb}^{-1}}}
\newc\nb{\ensuremath{\,\mathrm{nb}}} \newc\pb{\ensuremath{\,\mathrm{pb}}} \newc\fb{\ensuremath{\,\mathrm{fb}}}
\newc\pc{\ensuremath{\,\mathrm{pc}}}
\newc\kpc{\ensuremath{\,\mathrm{kpc}}}
\newc\mpc{\ensuremath{\,\mathrm{Mpc}}}
\newc\ps{\ensuremath{\,\mathrm{ps}}} 
\newc\cmeter{\ensuremath{\,\mathrm{cm}}} 
\newc\meter{\ensuremath{\,\mathrm{m}}} 
\newc\kmeter{\ensuremath{\,\mathrm{km}}}
\newc\second{\ensuremath{\,\mathrm{s}}}
\newc\msecond{\ensuremath{\,\mathrm{ms}}}
\newc\nsecond{\ensuremath{\,\mathrm{ns}}}
\newc\psecond{\ensuremath{\,\mathrm{ps}}}

\newc{\chisqmin}{\ensuremath{\chi^2_{\mathrm{min}}}}
\newc{\Delchisq}{\ensuremath{\Delta\chi^2}}
\newc{\chisq}{\ensuremath{\chi^2}}
\newc{\like}{\ensuremath{\mathcal{L}}}

\newc\lsim{\ensuremath{\mathrel{\rlap{\lower4pt\hbox{\hskip1pt$\sim$}}\raise1pt\hbox{$<$}}}}
\newc\gsim{\ensuremath{\mathrel{\rlap{\lower4pt\hbox{\hskip1pt$\sim$}}\raise1pt\hbox{$>$}}}}
\newc{\VEV}[1]{\ensuremath{\langle #1 \rangle}}
\newc{\dl}{\ensuremath{\stackrel{\leftarrow}{D}}}
\newc{\dr}{\ensuremath{\stackrel{\rightarrow}{D}}}

\newc{\bcenter}{\begin{center}}    \newc{\ecenter}{\end{center}}
\newc{\bfl}{\begin{flushleft}}    \newc{\efl}{\end{flushleft}}
\newc{\bfr}{\begin{flushright}}    \newc{\efr}{\end{flushright}}

\newc{\bi}{\begin{itemize}}
\newc{\ei}{\end{itemize}}
\newc{\bed}{\begin{description}}
\newc{\eed}{\end{description}}
\newc{\ben}{\begin{enumerate}}
\newc{\een}{\end{enumerate}}

\newc{\be}{\begin{equation}}
\newc{\ee}{\end{equation}}
\newc{\bea}{\begin{eqnarray}}
\newc{\eea}{\end{eqnarray}}
\newc{\bfle}{\begin{flalign}}
\newc{\efle}{\end{flalign}}
\newc{\ra}{\rightarrow}

\newc{\alphas}{\ensuremath{\alpha_s}}
\newc{\alphatwo}{\ensuremath{\alpha_2}}
\newc{\alphaone}{\ensuremath{\alpha_1}}
\newc{\alphai}[1]{\ensuremath{\alpha_{#1}}}
\newc{\alphaem}{\ensuremath{\alpha_{\mathrm{em}}}}
\newc{\alphaeff}{\ensuremath{\alpha_{\mathrm{eff}}}}
\newc{\sineff}{\ensuremath{\sin \theta_{\mathrm{eff}}}}
\newc{\sinsqeff}{\ensuremath{\sin^2 \theta_{\mathrm{eff}}}}
\newc{\dalphahad}{\ensuremath{\Delta \alpha_{\mathrm{had}}}}
\newc{\yt}{\ensuremath{h_t}} \newc{\yb}{\ensuremath{h_b}} \newc{\ytau}{\ensuremath{h_{\tau}}}
\newc\mz{\ensuremath{M_Z}} 
\newc\mw{\ensuremath{m_W}}
\newc\mZ{\mz}        \newc\mW{\mw}
\newc\mhsm{\ensuremath{ m_{H_{\mathrm{SM}}}}}
\newc{\mtop}{\ensuremath{ m_t}}               \newc{\mtpole}{\ensuremath{ M_t}}
\newc{\mbottom}{\ensuremath{ m_b}} 
\newc{\mtau}{\ensuremath{ m_{\tau}}}
\newc{\mt}{\mtpole}
\newc{\mb}{\mbottom} 
\newc{\rtwogg}{\ensuremath{R_{h_2}(\gamma\gamma)}}
\newc{\rtwozz}{\ensuremath{R_{h_2}(ZZ)}}
\newc{\ronegg}{\ensuremath{R_{h_1}(\gamma\gamma)}}
\newc{\ronezz}{\ensuremath{R_{h_1}(ZZ)}}
\newc{\rsiggg}{\ensuremath{R_{h_\textrm{sig}}(\gamma\gamma)}}
\newc{\rsigzz}{\ensuremath{R_{h_\textrm{sig}}(ZZ)}}
\newc{\llbar}{\ensuremath{\ell\bar{\ell}}}
\newc{\tauptaum}{\ensuremath{ \tau^+\tau^-}}
\newc{\qqbar}{\ensuremath{ q\bar{q}}} \newc{\ppbar}{\ensuremath{ p\bar{p}}}
\newc{\bbbar}{\ensuremath{ b\bar{b}}} \newc{\ttbar}{\ensuremath{ t\bar{t}}}
\newc{\ffbar}{\ensuremath{ f\bar{f}}} \newc{\tautaubar}{\ensuremath{ \tau\bar{\tau}}}

\newc{\mchi}{\ensuremath{m_\neutone}}
\newc{\squark}{\ensuremath{\tilde{q}}}
\newc{\slepton}{\ensuremath{\tilde{l}}}
\newc{\gluino}{\ensuremath{\tilde{g}}} 
\newc{\mgluino}{\ensuremath{{m_{\gluino}}}}
\newc{\wino}{\ensuremath{\tilde{W}}} 
\newc{\mwino}{\ensuremath{{m_{\wino}}}}
\newc{\tone}{\ensuremath{{\tilde{t}_1}}}
\newc{\bone}{\ensuremath{{\tilde{b}_1}}}
\newc{\Hone}{\ensuremath{{\tilde{H}_{1}}}}
\newc{\Htwo}{\ensuremath{{\tilde{H}_{2}}}}
\newc{\Hhtwo}{\ensuremath{{H_{2}}}}
\newc{\qli}{\ensuremath{{\tilde{Q}_{i}}}}
\newc{\uri}{\ensuremath{{\tilde{u}_{i}}}}
\newc{\dri}{\ensuremath{{\tilde{d}_{i}}}}
\newc{\lli}{\ensuremath{{\tilde{L}_{i}}}}
\newc{\eri}{\ensuremath{{\tilde{e}_{i}}}}

\newc{\sthw}{\ensuremath{ \sin\theta_W}}              \newc{\cthw}{\ensuremath{\cos\theta_W}}
\newc{\tanthw}{\ensuremath{ \tan\theta_W}}              \newc{\cotthw}{\ensuremath{\cot\theta_W}}
\newc{\ssqthw}{\ensuremath{\sin^2 \theta_W}}
\newc{\msbar}{\ensuremath{\overline{MS}}} \newc{\drbar}{\ensuremath{\overline{DR}}}
\newc{\mtmtsmmsbar}{\ensuremath{ m_t(m_t)^{\msbar}_{{\mathrm{SM}}}}}
\newc{\mtmtsmdrbar}{\ensuremath{ m_t(m_t)^{\drbar}_{{\mathrm{SM}}}}}
\newc{\mtmtmssmdrbar}{\ensuremath{ m_t(m_t)^{\drbar}_{{\mathrm{SUSY}}}}}
\newc{\mbmbmsbar}{\ensuremath{ m_b(m_b)^{\msbar} }}
\newc{\mbmbsmmsbar}{\ensuremath{ m_b(m_b)^{\msbar}_{{\mathrm{SM}}}}}
\newc{\mbmzsmmsbar}{\ensuremath{ m_b(\mz)^{\msbar}_{{\mathrm{SM}}}}}
\newc{\mbmzsmdrbar}{\ensuremath{ m_b(\mz)^{\drbar}_{{\mathrm{SM}}}}}
\newc{\mbmzmssmdrbar}{\ensuremath{ m_b(\mz)^{\drbar}_{{\mathrm{SUSY}}}}}
\newc{\mtaumzsmmsbar}{\ensuremath{ m_{\tau}(\mz)^{\msbar}_{{\mathrm{SM}}}}}
\newc{\mtaumzsmdrbar}{\ensuremath{ m_{\tau}(\mz)^{\drbar}_{{\mathrm{SM}}}}}
\newc{\mtaumzmssmdrbar}{\ensuremath{ m_{\tau}(\mz)^{\drbar}_{{\mathrm{SUSY}}}}}
\newc{\alphasmzms}{\ensuremath{\alpha_s(M_Z)^{\overline{MS}}}}
\newc{\alphaimzms}[1]{\ensuremath{\alpha_{#1}(M_Z)^{\overline{MS}}}}
\newc{\alphaemmz}{\ensuremath{\alpha_{\mathrm{em}}(M_Z)^{\overline{MS}}}}

\newc{\mzero}{\ensuremath{{m_0}}}
\newc{\mhalf}{\ensuremath{ m_{1/2}}}
\newc{\tanb}{\ensuremath{\tan\beta}}
\newc{\azero}{\ensuremath{ A_0}}
\newc{\signmu}{\ensuremath{\rm{sgn}\,\mu}}
\newc{\atau}{\ensuremath{{A_{\tau}}}}
\newc{\mueff}{\ensuremath{\mu_{\rm{eff}}}}
\newc{\lam}{\ensuremath{{\lambda}}}
\newc{\kap}{\ensuremath{{\kappa}}}
\newc{\alam}{\ensuremath{{A_{\lambda}}}}
\newc{\akap}{\ensuremath{{A_{\kappa}}}}
\newc{\hs}{\ensuremath{ H_s}}      
\newc{\mhs}{\ensuremath{ m_{H_s}}} 
\newc{\mgut}{\ensuremath{ M_{\rm GUT}}}
\newc{\gut}{\ensuremath{{\rm GUT}}}
\newc{\mplanck}{\ensuremath{ M_{\rm P}}}      \newc{\mpl}{\ensuremath{ M_{\rm Pl}}}
\newc{\msusy}{\ensuremath{ M_{\rm SUSY}}}      \newc{\ms}{\ensuremath{ M_{\rm S}}}
 \newc{\hu}{\ensuremath{ H_u}}       \newc{\hd}{\ensuremath{ H_d}}
 \newc{\mhu}{\ensuremath{ m_{H_u}}}       \newc{\mhd}{\ensuremath{ m_{H_d}}}
 \newc{\mhuew}{\ensuremath{ m^{\ast}_{H_u}}}       \newc{\mhdew}{\ensuremath{ m^{\ast}_{H_d}}}
 \newc{\mhuewsq}{\ensuremath{ m^{\ast\, 2}_{H_u}}}       \newc{\mhdewsq}{\ensuremath{ m^{\ast\, 2}_{H_d}}}
 \newc{\mhl}{\ensuremath{m_\hl}} 
 \newc{\mhone}{\ensuremath{m_{h_1}}} 
 \newc{\mhtwo}{\ensuremath{m_{h_2}}} 
 \newc{\mhi}{\ensuremath{m_{\tilde{h}}}} 
 \newc{\mul}{\ensuremath{m_{\tilde{u}_L}}} 
 \newc{\mbone}{\ensuremath{m_{\tilde{b}_1}}}  
 \newc{\mtone}{\ensuremath{m_{\tilde{t}_1}}} 
 \newc{\ma}{\ensuremath{m_A}} 
 \newc{\mH}{\ensuremath{m_H}} 
 \newc{\maone}{\ensuremath{m_{a_1}}} 
 \newc{\matwo}{\ensuremath{m_{a_2}}}
 \newc{\hone}{\ensuremath{h_1}}
 \newc{\htwo}{\ensuremath{h_2}}
 \newc{\aone}{\ensuremath{a_1}}
 \newc{\atwo}{\ensuremath{a_2}}
 \newc{\mqthree}{\ensuremath{m_{\tilde{Q}_3}^2}}
 \newc{\muthree}{\ensuremath{m_{\tilde{u}_3}^2}}
 \newc{\mqli}{\ensuremath{m_{\tilde{Q}_{i}}}}
 \newc{\muri}{\ensuremath{m_{\tilde{u}_{i}}}}
 \newc{\mdri}{\ensuremath{m_{\tilde{d}_{i}}}}
 \newc{\mlli}{\ensuremath{m_{\tilde{L}_{i}}}}
 \newc{\meri}{\ensuremath{m_{\tilde{e}_{i}}}}
 \newc{\ts}{\ensuremath{T_{SUSY}}}

\newc{\sigsip}{\ensuremath{\sigma^{\rm SI}_{p}}}	\newc{\sigsin}{\ensuremath{\sigma^{\rm SI}_{n}}}
\newc{\sigsdp}{\ensuremath{\sigma^{\rm SD}_{p}}}	\newc{\sigsdn}{\ensuremath{\sigma^{\rm SD}_{n}}}
\newc{\sigsi}{\ensuremath{\sigma^{\rm SI}}}	\newc{\sigsd}{\ensuremath{\sigma^{\rm SD}}}
\newc{\abund}{\ensuremath{ \Omega h^2}}
\newc{\omegadm}{\ensuremath{ \Omega_{{\rm DM}}}}     \newc{\abunddm}{\ensuremath{ \Omega_{{\rm DM}} h^2}} 
\newc{\omegam}{\ensuremath{ \Omega_{{\rm m}}}}       \newc{\abundm}{\ensuremath{ \Omega_{{\rm m}} h^2}}
\newc{\omegab}{\ensuremath{ \Omega_{{\rm b}}}}	\newc{\abundb}{\ensuremath{ \Omega_{{\rm b}} h^2}}
\newc{\omegatot}{\ensuremath{ \Omega_{{\rm TOT}}}}
\newc{\omegacdm}{\ensuremath{ \Omega_{{\rm CDM}}}}   \newc{\abundcdm}{\ensuremath{ \Omega_{{\rm CDM}} h^2}}
\newc{\omegalambda}{\ensuremath{ \Omega_{\Lambda}}} \newc{\abundlambda}{\ensuremath{ \Omega_{\Lambda} h^2}}
\newc{\omegarad}{\ensuremath{ \Omega_{{\rm rad}}}}  \newc{\abundrad}{\ensuremath{ \Omega_{{\rm rad}} h^2}}
\newc{\rhocrit}{\ensuremath{ \rho_{\rm crit}}}
\newc{\rhochi}{\ensuremath{ \rho_{\chi}}}
\newc{\abunchi}{\ensuremath{\Omega_\chi h^2}}
\newc{\abundlsp}{\ensuremath{\Omega_{\rm LSP}h^2}}

\newc{\amu}{\ensuremath{ a_{\mu}}}        \newc{\amususy}{\ensuremath{ a_{\mu}^{\mathrm{SUSY}}}}
\newc{\amuexpt}{\ensuremath{ a_{\mu}^{\mathrm{expt}}}}        \newc{\amusm}{\ensuremath{ a_{\mu}^{\mathrm{SM}}}}
\newc\deltaamu{\ensuremath{\Delta a_{\mu}}} \newc{\deltaamususy}{\ensuremath{\delta a_{\mu}^{\mathrm{SUSY}}}}
\newc\gmtwo{\ensuremath{ (g-2)_{\mu}}} 
\newc{\deltagmtwomususy}{\ensuremath{\delta\left(g-2\right)_{\mu}^{\mathrm{SUSY}}}}
\newc{\deltagmtwomu}{\ensuremath{\delta\left(g-2\right)_{\mu}}}
\newc\BR{\ensuremath{\rm BR}}
\newc\bsgamma{\ensuremath{ b\rightarrow s \gamma }}
\newc\bxsgamma{\ensuremath{\overline{B}\rightarrow X_{s}\gamma}}
\newc\brbsgamma{\ensuremath{\BR\left(\bsgamma\right)}}
\newc\brbxsgamma{\ensuremath{\BR\left(\bxsgamma\right)}}
\newc\bsmumu{\ensuremath{B_s\to\mu^+\mu^-}}
\newc\brbsmumu{\ensuremath{\BR\left(B_s\to\mu^+\mu^-\right)}}
\newc\bdmmumu{\ensuremath{\overline{B}_d\to\mu^+\mu^-}}
\newc\bbbarmix{\ensuremath{\overline{B}_s\mbox{-}B_s}}      
\newc\delmbs{\ensuremath{\Delta M_{B_s}}}
\newc{\butaunu}{\ensuremath{B_u \rightarrow \tau \nu}}
\newc{\brbutaunu}{\ensuremath{\BR\left(B_u \rightarrow \tau \nu\right)}}

\newcommand*{\reffig}[1]{Fig.~\ref{#1}}
 
        \newcommand*{\refeq}[1]{Eq.~(\ref{#1})}
     \newcommand*{\refsec}[1]{Sec.~\ref{#1}}

\newcommand*{\neutone}{\ensuremath{\chi^0_1}}

\let\oldcite\cite
\renewcommand*{\cite}{~\oldcite}

\newcommand*{\hl}{\ensuremath{h}}

\restylefloat{figure}

\title{Gauge contribution to the $1/N_F$ expansion of the Yukawa coupling beta function}

\author{Kamila Kowalska}
\author{and Enrico Maria Sessolo}

\affiliation{National Centre for Nuclear Research,\\
Ho{\.z}a 69, 00-681 Warsaw, Poland} 

\emailAdd{kamila.kowalska@ncbj.gov.pl}
\emailAdd{enrico.sessolo@ncbj.gov.pl}

\abstract{We provide a closed analytical form for the gauge contribution to the beta function of a generic Yukawa coupling in the limit of large $N_F$, where $N_F$ is the number of heavy vector-like fermions charged under an abelian or non-abelian gauge group. The resummed expression is finite and for the abelian case presents a pole at the same location as for the corresponding gauge beta function. 
When applied to new physics scenarios characterized by large Yukawa couplings, the contribution calculated here can cure their pathological 
UV behavior and make the couplings asymptotically free.}

\begin{document}
\maketitle

\section{Introduction\label{sec:intro}}

In light of the lack of clear signals for new physics at the Large Hadron Collider (LHC), alternatives to the most popular extensions
of the Standard Model (SM) have experienced a recent revival. Asymptotic safety\cite{Wilson:1971bg,Weinberg:1980gg} is one of the ideas that have been brought back, to be adapted, for example, to the framework of four-dimensional gauge theories\cite{Litim:2014uca}. 
In the context of asymptotic safety, very recently renewed attention has been paid\cite{Mann:2017wzh,Abel:2017rwl,Pelaggi:2017abg,Antipin:2017ebo} to the large-$N_F$ expansion of the gauge coupling beta function, which was calculated in closed analytical form a few decades ago\cite{PalanquesMestre:1983zy,Gracey:1996he} (see also\cite{Espriu:1982pb}). Assuming the presence of $N_F$ vector-like heavy fermions charged under an abelian 
or non-abelian group with coupling $\alpha$, corrections to the gauge boson propagator by chain-diagrams with an ever increasing 
number of vacuum-polarization bubbles can be systematically collected in a power series of $K=\alpha N_F/\pi$. 
At the leading order in $1/N_F$ the momentum-independent series admits an analytical limit in the $\overline{MS}$ scheme, 
with a finite radius of convergence, thus providing a closed expression for the beta function when $N_F\rightarrow\infty$ 
(see\cite{Holdom:2010qs} for a review). 

The fact that the analytical form of the gauge beta function constructed in this way presents a negative pole at $K=15/2$ has been used
in\cite{Holdom:2010qs,Mann:2017wzh} to remove the Landau pole in the running of the SM U(1) gauge coupling, 
and thus give rise to an asymptotically safe extension characterized by an interacting UV fixed point. 
On the other hand, as is mentioned in passing in\cite{Mann:2017wzh}, but not pursued, one also expects an impact on the beta 
function of eventually present Yukawa couplings, as these will be affected by the same chain-diagram $N_F$ fermion contributions to the gauge boson propagator.  

In particular, since one observes\cite{PalanquesMestre:1983zy,Antipin:2017ebo} 
that the anomalous dimension of the fermion mass presents a pole in the same position as for
the U(1) gauge beta function, because they are similarly affected by the $N_F$ resummation,  
one should also expect the presence of a pole at $K=15/2$ for the Yukawa coupling beta function, since 
the same chain contributions appear in the renormalization of the Yukawa vertex. 
Note, however, that it is important to map correctly 
the analytical behavior of the corresponding beta function close to the singularity to actually determine whether 
the full theory could be considered safe around the gauge coupling fixed point.
  
In this paper we fill a gap in the relative literature and 
calculate the beta function of a generic Yukawa coupling at order $1/N_F$ in the presence of $N_F$ heavy 
vector-like pairs of fermions, all of which are charged under the gauge group, but of which only a small subset couples to the scalar field. We provide a closed form solution similar to the one derived for the gauge couplings in\cite{PalanquesMestre:1983zy,Gracey:1996he} 
and illustrate 
the effects of this calculation on the running of the SM and new physics Yukawa couplings, which remain asymptotically free for 
a broad range of low-scale boundary conditions.

We point out that this result could have interesting phenomenological consequences, 
in particular in scenarios where sizable Yukawa couplings between the SM and an extra sector
are required to accommodate some experimental anomalies.
(Recent cases of this type include, e.g., loop-induced explanations of the LHCb flavor anomalies\cite{Belanger:2015nma,Gripaios:2015gra,Arnan:2016cpy,DAmico:2017mtc,Kawamura:2017ecz}, or
some of the proposed explanations for the muon $g-2$ discrepancy\cite{Freitas:2014pua,Kowalska:2017iqv}.)
In phenomenological models of this kind the fact that new Yukawa couplings usually become non-perturbative for scales as low as 
few hundred~TeV is often considered a symptom of pathology. This does not need to be the case 
if the renormalization group (RG) evolution of the new couplings remains under control.
It is important to be aware, however, that the full asymptotic safety of the theory is not yet at this level guaranteed, as one should then proceed to 
calculate the resummed contributions to scalar quartic coupling. 

The structure of the paper is as follows. In \refsec{sec:calc} we outline the calculation techniques and obtain 
the Yukawa coupling beta function in the limit of large $N_F$. In \refsec{sec:properties} we discuss the properties of the solution and show its impact on the running of the Yukawa couplings. We summarize our findings in \refsec{sec:summary}. 

\section{The large-$N_F$ expansion of beta functions\label{sec:calc}}

We start this section by briefly reviewing the form of the large-$N_F$ expansion of the U(1) gauge beta function, 
which was first derived in\cite{PalanquesMestre:1983zy}. 
This allows us to set the tone for the Yukawa coupling calculation and introduce 
the notation we use throughout the paper. 

\subsection{Abelian gauge coupling}\label{sec:gauge}

Consider a generic quantum field theory with an abelian gauge symmetry U(1), whose interaction strength is given by $\alpha_1=g_1^2/4\pi$, and assume that the theory contains $N_F$ vector-like fermions, whose charge under the gauge symmetry is given by $q$. 
It is formally possible to expand the perturbation theory in the parameter $K=\alpha_1 q^2 N_F/\pi$, which, for the remainder of this section, will play the role of a rescaled coupling constant kept fixed in the limit of large $N_F$. 

In $d=4-\epsilon$ space-time dimensions the renormalized coupling constant is then related to the bare coupling constant $K_0$ by 
$K=Z_3\,\mu^{-\epsilon}K_0$, where $\mu$ is the renormalization scale. 
The renormalization constant $Z_3$, can then be formally expanded at all orders in $1/N_F$ as
\be
Z_3=1-\frac{2}{3}\frac{K}{\epsilon}-\frac{1}{N_F}\sum_{n=2}^{\infty}K^n\sum_{i=0}^{n-1}\frac{B_i^{(n)}}{\epsilon^{n-i}}
-\frac{1}{N_F^2}\sum_{n=3}^{\infty}K^n\sum_{i=0}^{n-1}\frac{C_i^{(n)}}{\epsilon^{n-i}}+...\label{gaugeZ3exp}\,,
\ee  
in terms of appropriate coefficients $B_i^{(n)}, C_i^{(n)}$. Ideally, for sufficiently large $N_F$ 
one just needs to retain the first order in $1/N_F$.

It is straightforward to show\cite{PalanquesMestre:1983zy} that the gauge beta function equivalently admits the expansion   
\bea
\beta_1(K)\equiv \frac{d \log K}{d \log\mu}&=&\frac{2}{3}K+\frac{1}{N_F}\sum_{n=2}^{\infty}nK^n B^{(n)}_{n-1}+...\nonumber\\
 &=&\frac{2}{3}K\left(1+\sum_{i=1}^{\infty}\frac{F_i(K)}{N_F^i}\right)\,,\label{gaugebetaexp}
\eea
at all orders in perturbation theory. Again, in the limit of large $N_F$, one would just like to retain the dominant contribution 
in $1/N_F$, which is entirely parameterized by the function $F_1(K)$. 

\begin{figure}[t]
\centering
\includegraphics[width=0.3\textwidth]{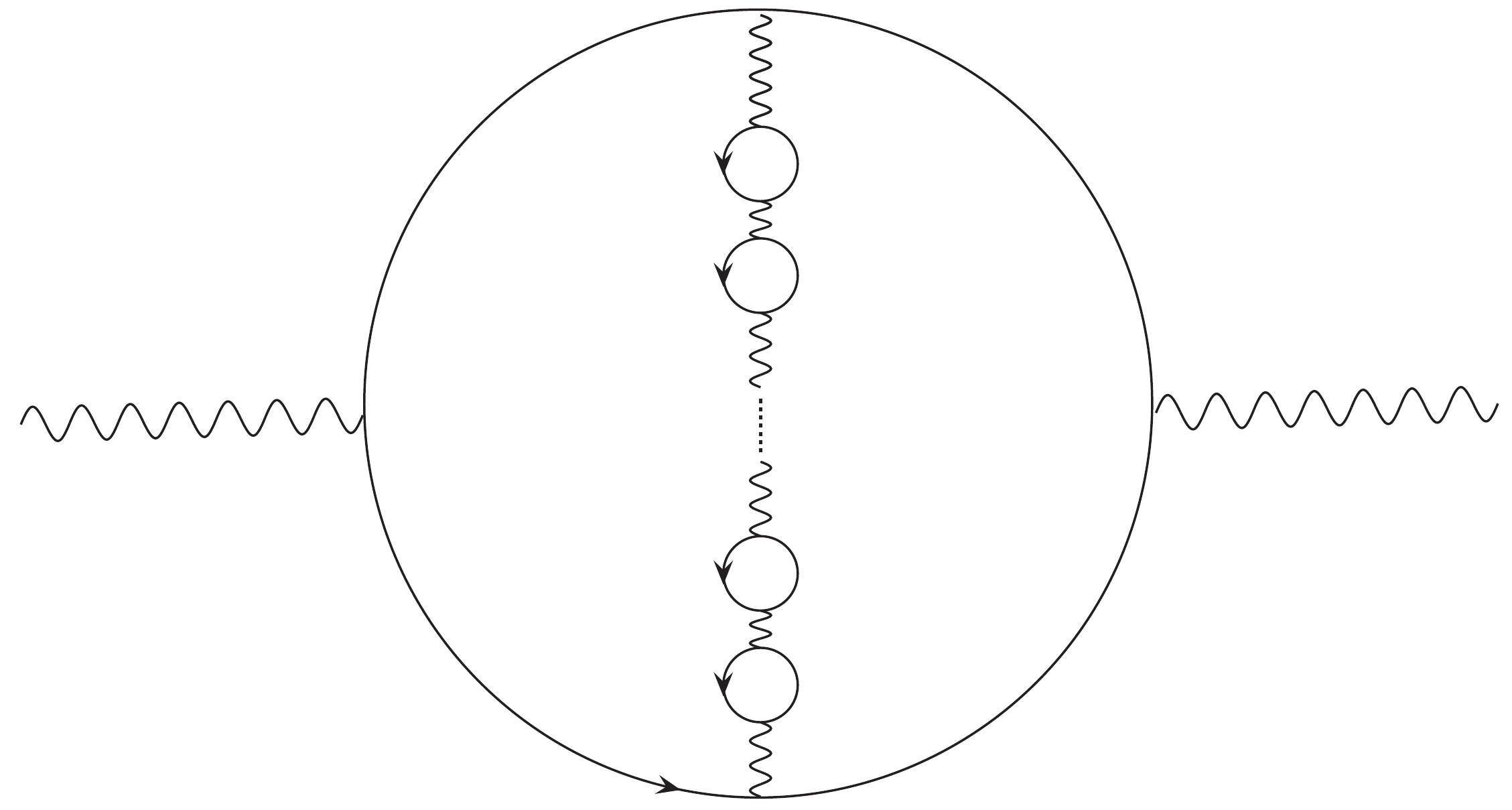}
\caption{An example of loop diagram contributions to the large-$N_F$ expansion of the gauge boson vacuum polarization 
at order $1/N_F$. Each bubble produces a term proportional to $K\sim \alpha_1 N_F$.}
\label{fig:feyn_Gau}
\end{figure}

The function $F_1(K)$ emerges from resummation of the infinite number of loop contributions to the gauge boson self-energy, 
an example of which is depicted in \reffig{fig:feyn_Gau}. 
Reference\cite{PalanquesMestre:1983zy} first showed that this admits a closed analytical form.
For Q.E.D. it is given by 
\be\label{ef1}
F_1(K)=
\int_0^{K/3}dx\,\frac{\left(1+x\right)\left(2x-1\right)^2\left(2x-3\right)^3\sin\left(\pi x\right)^3\Gamma\left(x-1\right)^2\Gamma\left(-2x\right)}{\left(x-2\right)\pi^3}\,.
\ee
Closed forms for the $1/N_F$ term of the large-$N_F$ expansion have also been calculated for the beta function 
of Q.C.D.\cite{Gracey:1996he}, and for the renormalization group $\gamma$ function in the abelian and non-abelian case\cite{PalanquesMestre:1983zy,Espriu:1982pb}.\footnote{The anomalous dimension of the 3-point vertex, 
in a theory with U(1)-charged fermions and a Yukawa interaction, was calculated in finite form at orders $1/N_F$, $1/N_F^2$ 
in\cite{Gracey:1991ry,Gracey:1993zn,Gracey:1993ai} using the self-consistency technique of critical exponents.}  

It is important to point out that each fermion loop introduces an enhancement factor $N_F$, so that the validity of the perturbation theory expansion in $K\sim \alpha_1 N_F$ can be called into question and the convergence of the infinite series of loop-diagrams must be checked. It was pointed out in\cite{Holdom:2010qs} that the $1/N_F$ expansion of \refeq{gaugebetaexp} 
is under control if $N_F\geq 16$, as order by order in $K$ the term in $1/N_F^{n+1}$ becomes suppressed 
with respect to the corresponding term in $1/N_F^{n}$. This requirement provides a lower bound on the allowed number of vector-like fermions.


\subsection{Yukawa coupling}\label{yukabel}

We now extend the gauge theory introduced in \refsec{sec:gauge} 
by a generic Yukawa interaction between a scalar field $\phi$ and a pair of fermions $\psi_L$, $\psi_R$\,:
\be\label{yukint}
\mathcal{L}_{\textrm{Yuk}}=-y\,\bar{\psi}_L\,\phi\,\psi_R+\textrm{h.c.}
\ee 
All the participating fields can be charged under the U(1), and we denote the respective charges as $q_S$, $q_L$, and $q_R$. 
We also define $\alpha_y=y^2/4\pi$. Fermions $\psi_L, \psi_R$ may either belong to the set of $N_F$ vector-like fermions, 
or they may simply represent additional degrees of freedom, for example quarks and leptons of the Standard Model. 

From the above setting one can easily see that the self-energy corrections to the scalar field 
are not enhanced by bubbles of $N_F$ fermions, unlike the vacuum polarization of the gauge boson, as only one pair of fermions couples to the scalar field. The latter assumption thus considerably simplifies the calculation of the Yukawa beta function 
and we adopt it throughout this paper. Whether relaxing this assumption may still lead to a 
closed analytical form for the Yukawa beta function 
remains a non-trivial and open question, and we reserve it for future investigation. 

In a fashion analogous to \refeq{gaugebetaexp}, one can formally expand the Yukawa beta function as 
\be\label{yukNF}
\beta_y(K)\equiv\frac{d\ln\alpha_y}{d\ln\mu}=\sum_{i=1}^{\infty}\frac{Y_i(K)}{N_F^i}\,,
\ee
where we neglect to explicitly write additional finite terms that at one loop include 
the standard fermion-scalar-fermion vertex correction and, if relevant, contributions from other gauge groups.
Our goal is to derive a closed analytical form for the first-order coefficient $Y_1(K)$ of the expansion~(\ref{yukNF}). 
As pertains to higher orders in $1/N_F$, the $K^2$ term of $Y_2(K)$ was derived in Ref.\cite{Machacek:1983fi}. 
We have checked that for $N_F\geq 16$ it is much smaller than the $K^2$ term of $Y_1(K)$, see the end of \refsec{sec:nonabel} for its explicit form.
We assume here that all other higher-order terms are negligible for $N_F\geq 16$.

\begin{figure}[t]
\centering
\includegraphics[width=0.21\textwidth]{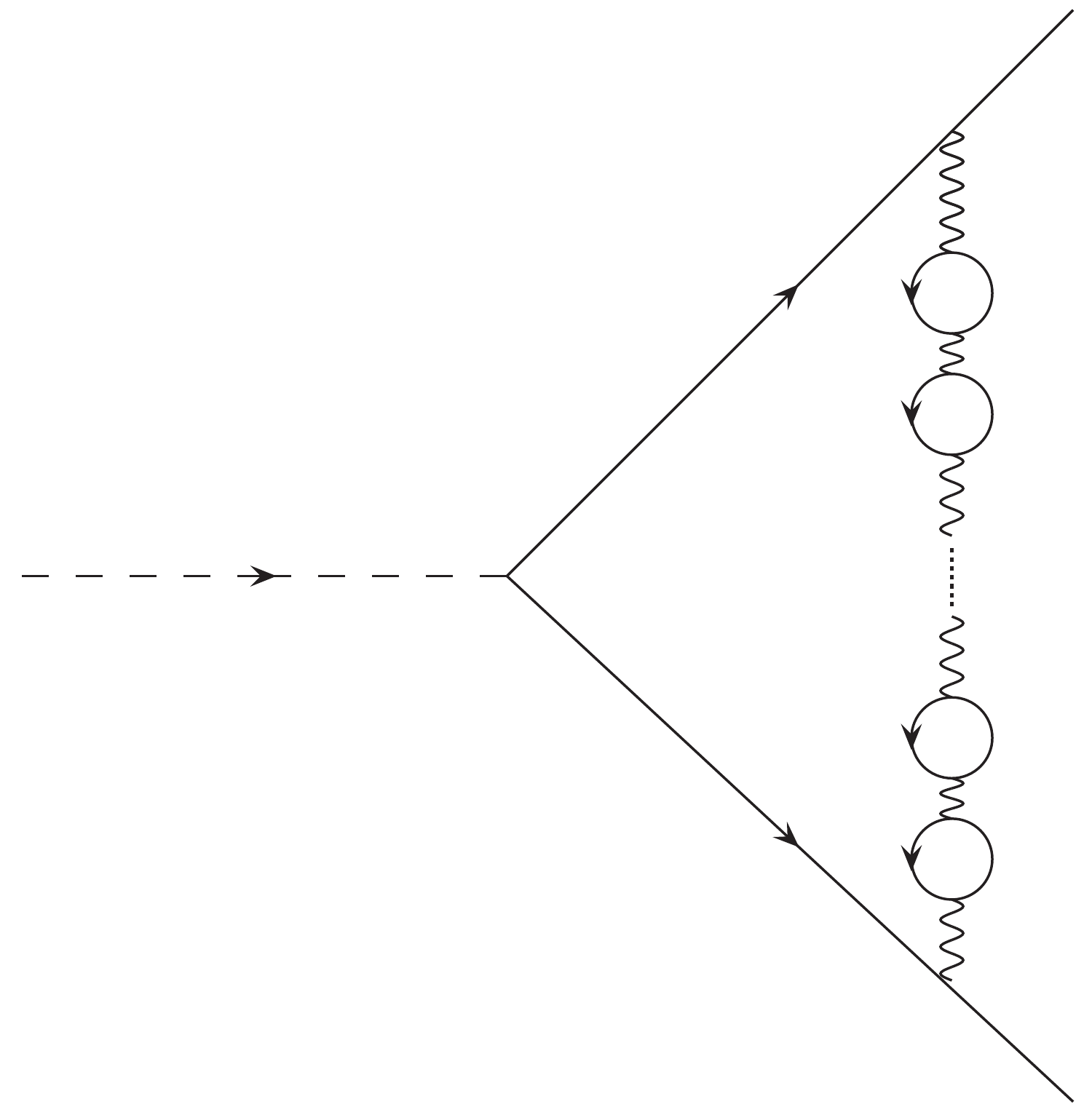}
\hspace{0.5cm}
\includegraphics[width=0.2\textwidth]{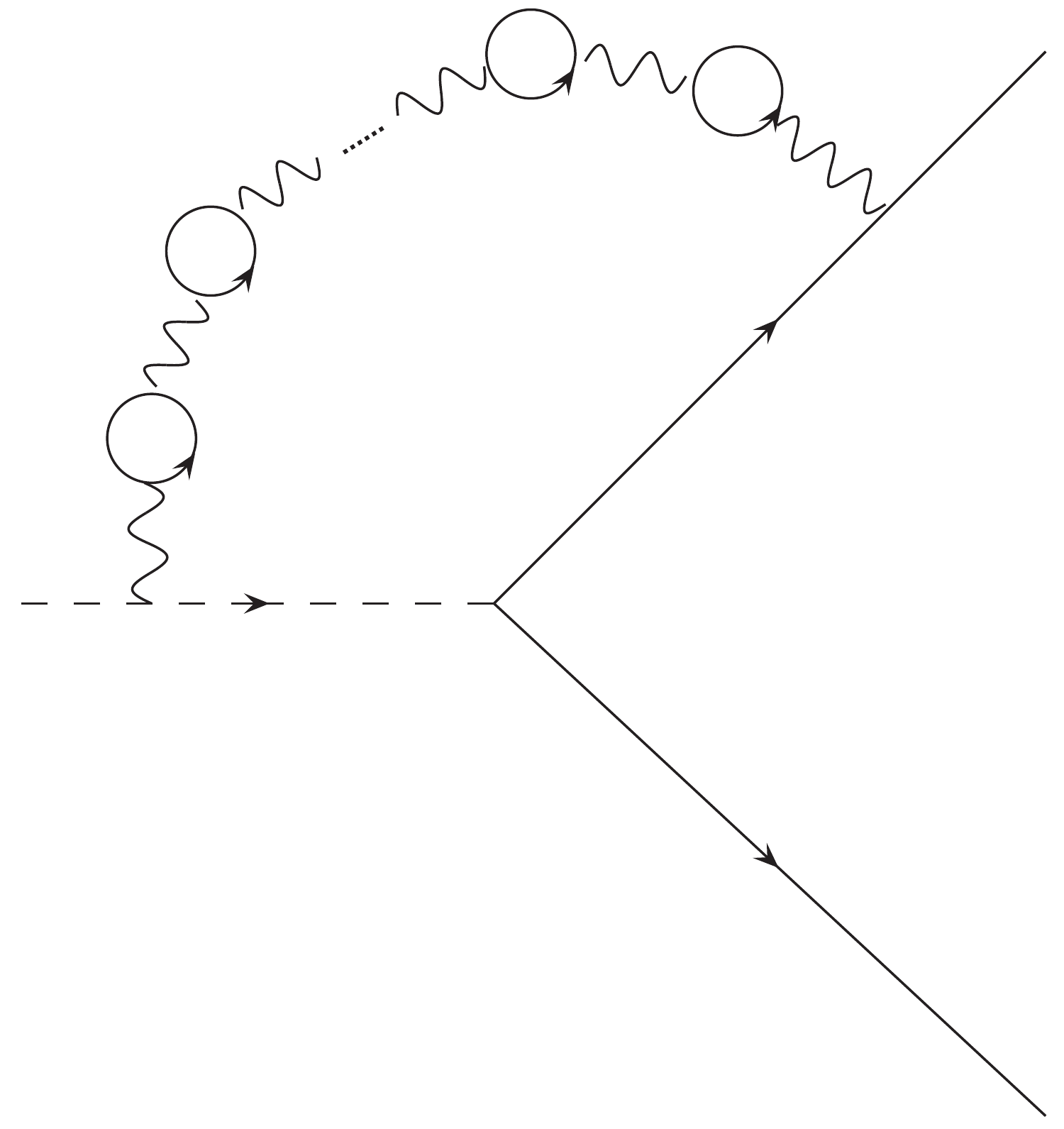}
\hspace{0.5cm}
\includegraphics[width=0.2\textwidth]{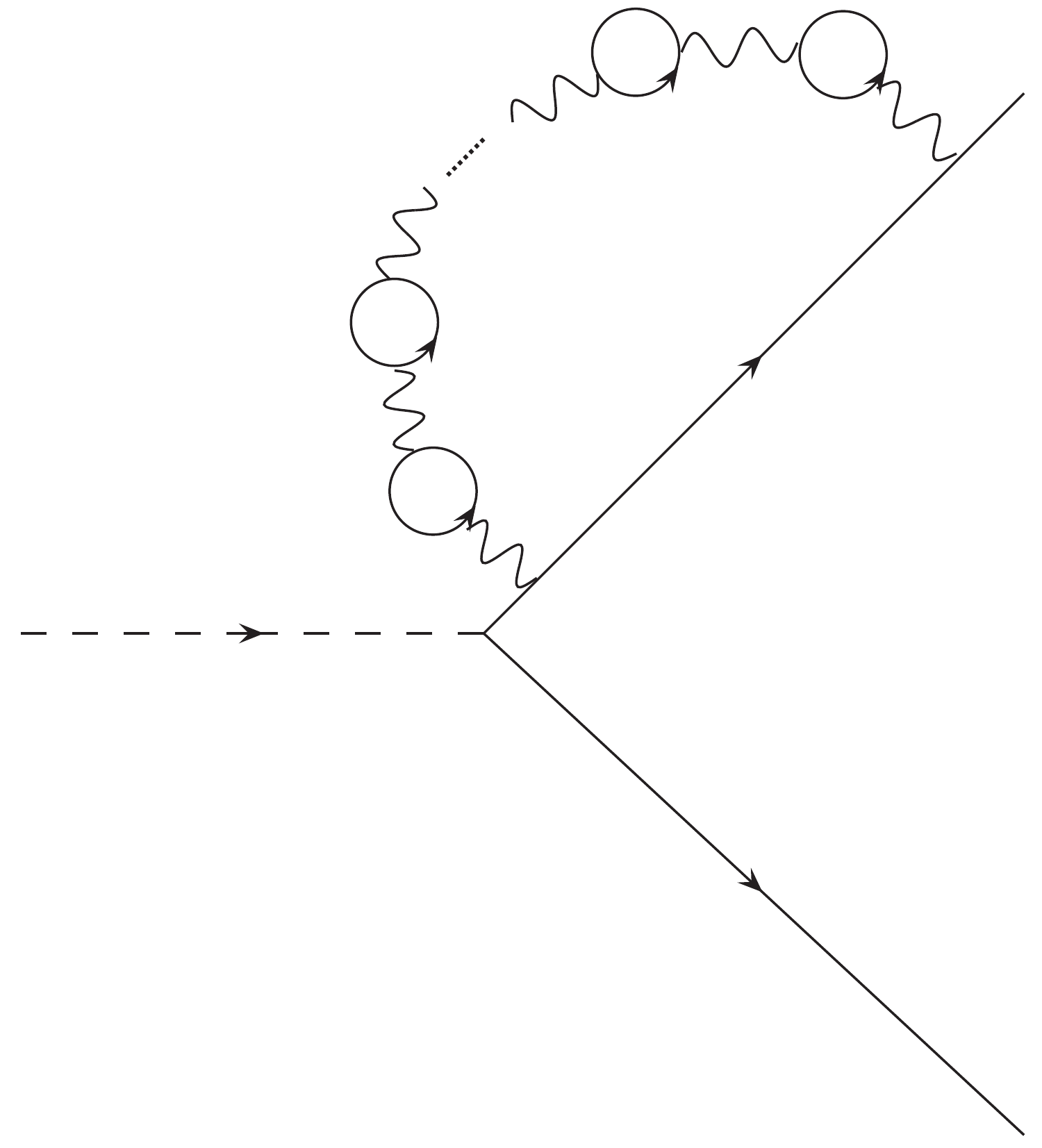}
\hspace{0.5cm}
\includegraphics[width=0.2\textwidth]{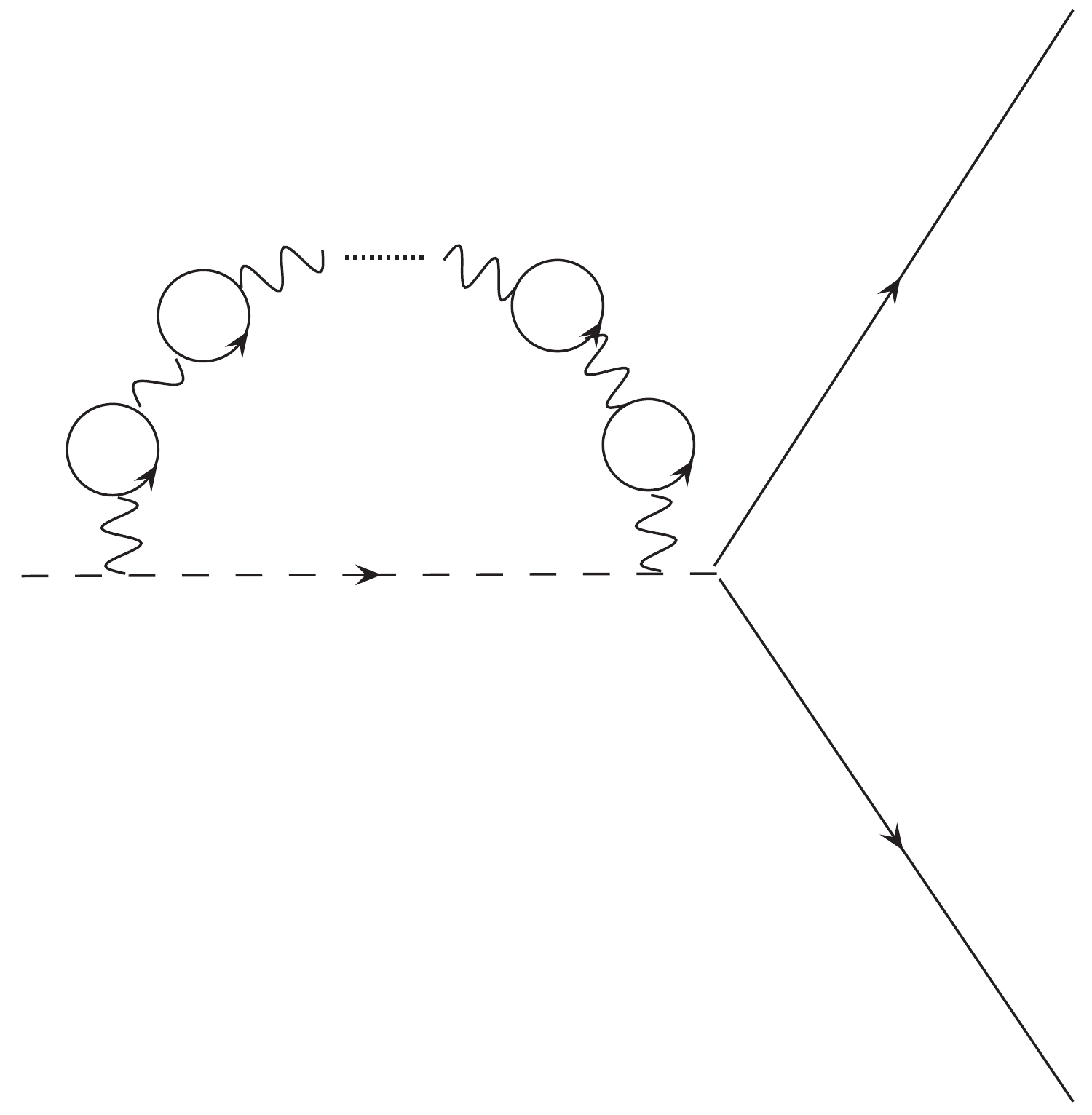}
\caption{Dominant contributions to the Yukawa vertex in the limit of large $N_F$.}
\label{feyn_Yuk}
\end{figure}

In order to derive the gauge contribution to $\beta_y$, one needs to calculate the loop diagrams shown in \reffig{feyn_Yuk}. 
We adopt in this paper the $\overline{MS}$ renormalization scheme. We also choose the Feynman gauge for the gauge boson propagator, 
as in the $\overline{MS}$ scheme the beta functions are gauge-parameter independent\cite{Ozer:1998ym}. 

In terms of renormalized quantities, the Lagrangian of the theory is given by
\be
\mathcal{L}\supset iZ_{2\,L}\bar{\psi}_L\slashed{\partial}\psi_L+iZ_{2\,R}\bar{\psi}_R\slashed{\partial}\psi_R+\frac{1}{2}Z_\phi(\partial_\mu\phi)^2-y\,Z_1\bar{\psi}_L\phi\,\psi_R+\textrm{h.c.}\,,
\ee
where we neglect to explicitly write the mass terms. 
In $d=4-\epsilon$ dimensions the renormalization constants, which relate renormalized and bare quantities 
(the latter indicated in what follows by the subscript $0$), are defined so that
\bea
\psi_{L,0}&=&\left(Z_{2\,L}\,\mu^{-\epsilon}\right)^{1/2}\psi_L\,,\\
\psi_{R,0}&=&\left(Z_{2\,R}\,\mu^{-\epsilon}\right)^{1/2}\psi_R\,,\\
\phi_0&=&\left(Z_{\phi}\,\mu^{-\epsilon}\right)^{1/2}\phi\,,\\
\alpha_{y,0}&=&Z_{\phi}^{-1}\,Z_{2\,L}^{-1}\,Z_{2\,R}^{-1}\,Z_1^2\,\mu^{\epsilon}\,\alpha_y\,.\label{bare_yuk}
\eea

We define the counterterms needed to cancel the divergent parts of the loop contributions to the fermion self-energies, $\Sigma_{2}(\slashed{p})_{L,R}$\,, 
the scalar self-energy, $S(p^2)$, and the Yukawa coupling vertex function, $\Lambda_Y(p,p')$,
as 
\be\label{rencon}
Z_{2\,L,R}-1=\frac{d}{d\slashed{p}}\Sigma_2(\slashed{p})_{L,R}\Big|_{\textrm{poles in }\epsilon}\quad Z_\phi-1=\frac{d}{d p^2}S(p^2)\Big|_{\textrm{poles in }\epsilon}\quad Z_1-1=-\Lambda_Y(p,p')\Big|_{\textrm{poles in }\epsilon}\,,
\ee
and the corresponding quantities, which emerge from resumming an infinite series of fermion loops as in \reffig{feyn_Yuk}, are given by
\bea\label{summed}
-i\Sigma_2(\slashed{p})_{L,R}=\sum_{n=0}^{\infty}\left[-i\Sigma_2^{(n)}(\slashed{p})_{L,R}\right]\,,\quad -iS(p^2)=\sum_{n=0}^{\infty}\left[-iS^{(n)}(p^2)\right]\,,\nonumber\\
\Lambda_Y(p,p')=\sum_{n=0}^{\infty}\Lambda_Y^{(n)}(p,p')\,,\qquad\qquad\qquad
\eea
where it is understood that the amplitude for the general self-energy diagrams of the Dirac-type fermions 
is given by $-i\Sigma_2=-i\left(\Sigma_{2\,L} P_L+\Sigma_{2\,R} P_R\right)$.
We repeat that in (\ref{summed}) we do not explicitly consider 
direct contributions from fermion bubbles to the scalar propagator, which are well known and not modified by $N_F$ in our setup.

At the $n$-th order, the quantities of \refeq{summed} are then obtained by calculating the following integrals (see also 
Appendix~\ref{appen:vac}):
\be
-i\Sigma_2^{(n)}\left(\slashed{p}\right)_{L,R}=\left(-i g_{1,0}\,q_{L,R}\right)^2\,\mu^\epsilon\,\Pi^n(0)\int\frac{d^d k}{(2\pi)^d}\gamma_{\mu}\frac{i\left(\slashed{p}-\slashed{k}+m_0\right)}{\left(p-k\right)^2-m_0^2}\gamma^{\mu} 
\frac{-i\left(\mu^2\right)^{\frac{n\epsilon}{2}}}{\left(k^2\right)^{1+\frac{n\epsilon}{2}}}\,,\label{fermion}
\ee
\be
-iS^{(n)}\left(p^2\right)=(-i g_{1,0}\,q_S)^2\,\mu^\epsilon\,\Pi^n(0)\int\frac{d^d k}{(2\pi)^d}\frac{i\left(2p-k\right)_\mu
\left(2p-k\right)^\mu}{(p-k)^2-M_0^2}\,
\frac{-i\left(\mu^2\right)^{\frac{n\epsilon}{2}}}{\left(k^2\right)^{1+\frac{n\epsilon}{2}}}\,,\label{sca_self}
\ee
\begin{multline}
\Lambda_Y^{(n)}\left(p,p'\right)=\left(-i g_{1,0}\right)^2 q_L q_R\,\mu^{\epsilon}\,\Pi^n(0)\int\frac{d^d k}{(2\pi)^d} \gamma_{\nu}\frac{i\left(\slashed{p}-\slashed{k}+m_0\right)}{\left(p-k\right)^2-m_0^2}y\frac{i\left(\slashed{p}'-\slashed{k}+m_0\right)}{\left(p'-k\right)^2-m_0^2}\gamma^{\nu}\frac{-i\left(\mu^2\right)^{\frac{n\epsilon}{2}}}{\left(k^2\right)^{1+\frac{n\epsilon}{2}}}\\
+\,q_S^2\left(-i g_{1,0}\right)^2\mu^{\epsilon}\,\Pi^n(0)\int\frac{d^d k}{(2\pi)^d} \gamma_{\nu}\frac{i\left(\slashed{p}'-\slashed{k}+m_0\right)}{\left(p'-k\right)^2-m_0^2}y\frac{i\left(2 l-k\right)^{\nu}}{\left(l-k\right)^2-M_0^2}\frac{-i\left(\mu^2\right)^{\frac{n\epsilon}{2}}}{\left(k^2\right)^{1+\frac{n\epsilon}{2}}}\,,\label{vertex}
\end{multline}
where $l=p'-p$, $m_0$ is the mass of the fermions $\psi_{L,R}$\,, and $M_0$ is the mass of the scalar field $\phi$.
In Eqs.~(\ref{fermion})-(\ref{vertex}), $\Pi(0)$ gives the momentum-independent part of the 1PI one-loop contribution to the photon propagator (see Appendix~\ref{appen:vac}):
\be\label{pizen}
\Pi(k^2)=\left(\frac{\mu^2}{k^2}\right)^{\frac{\epsilon}{2}}\Pi(0)=-2 K_0\left(-\frac{4\pi\mu^2}{k^2}\right)^{\frac{\epsilon}{2}}\frac{\Gamma\left(\frac{\epsilon}{2}\right)\Gamma\left(2-\frac{\epsilon}{2}\right)^2}{\Gamma\left(4-\epsilon\right)}\,.
\ee

Since the $\overline{MS}$ scheme is mass-independent, we can take the limit $m_0^2,M_0^2\ll p^2$ in 
Eqs.~(\ref{fermion})-(\ref{sca_self}) to simplify the calculation. For the same reason, we work in the limit 
$p=p'=0$ to evaluate \refeq{vertex}, since its divergent part does not depend on the external momenta. 
The integrals can thus be easily reduced using standard techniques. 

By performing the substitution $g_{1,0}^2\rightarrow 4\pi^2 K_0/(N_Fq^2)$, and rescaling $n\rightarrow n-1$, 
one can arrive to the final $1/N_F$ expansions of Eqs.~(\ref{rencon})-(\ref{summed}),
\bea
\frac{d\Sigma_2(\slashed{p})_{L,R}}{d\slashed{p}}&=&\frac{3}{4N_F}\frac{q^2_{L,R}}{q^2}\sum_{n=1}^\infty\left(-\frac{2}{3}K_0\right)^n\frac{1}{n\epsilon ^n}H_2(n,\epsilon)\,,\label{loops}\\
\frac{dS(p^2)}{dp^2}&=&-\frac{3}{2N_F}\frac{q^2_S}{q^2}\sum_{n=1}^\infty\left(-\frac{2}{3}K_0\right)^n\frac{1}{n\epsilon ^n}H_\phi(n,\epsilon)\,,\label{loopsS}\\
\Lambda_Y(0,0)&=&-\frac{3y}{N_F}\sum_{n=1}^{\infty}\left(-\frac{2}{3}K_0\right)^n\frac{1}{n\epsilon^n}\left[\frac{q_Lq_R}{q^2}Y_1\left(n,\epsilon\right)+\frac{1}{4}\frac{q_S^2}{q^2}Y_2\left(n,\epsilon\right)\right]\,,\label{loopsL}
\eea
where the functions $H_2(n,\epsilon)$, $H_\phi(n,\epsilon)$, $Y_1(n,\epsilon)$, and $Y_2(n,\epsilon)$ are finite and given by 
\begin{multline}\label{poly1}
H_2(n,\epsilon)=\left(-\frac{4\pi\mu^2}{p^2}\right)^{\frac{n\epsilon}{2}}\left[\frac{\Gamma\left(1+\frac{\epsilon}{2}\right)\Gamma\left(1-\frac{\epsilon}{2}\right)^2\left(1-\frac{\epsilon}{2}\right)}{\Gamma\left(1-\epsilon\right)\left(1-\epsilon\right)\left(1-\frac{\epsilon}{3}\right)}\right]^n\frac{\Gamma\left(1-\epsilon\right)}{\Gamma\left(1+\frac{\epsilon}{2}\right)\Gamma\left(1-\frac{\epsilon}{2}\right)}\\
\times\,\frac{\Gamma\left(1+\frac{n\epsilon}{2}\right)\Gamma\left(1-\frac{n\epsilon}{2}\right)}{\Gamma\left(1+\frac{n\epsilon}{2}-\frac{\epsilon}{2}\right)\Gamma\left(1-\frac{n\epsilon}{2}-\frac{\epsilon}{2}\right)}\frac{\left(1-\epsilon\right)\left(1-\frac{\epsilon}{3}\right)\left[1-\epsilon\left(1-3n\right)+\frac{\epsilon^2}{4}\left(1-4n-4n^2\right)+\frac{n^2\epsilon^3}{4}\right]}{\left(1-\frac{\epsilon}{2}\right)\left(1-\frac{\epsilon}{4}-\frac{n\epsilon}{4}\right)\left(1-\frac{\epsilon}{2}-\frac{n\epsilon}{2}\right)}\,,
\end{multline}
\begin{multline}\label{poly2}
H_\phi(n,\epsilon)=\left(-\frac{4\pi\mu^2}{p^2}\right)^{\frac{n\epsilon}{2}}\left[\frac{\Gamma\left(1+\frac{\epsilon}{2}\right)\Gamma\left(1-\frac{\epsilon}{2}\right)^2\left(1-\frac{\epsilon}{2}\right)}{\Gamma\left(1-\epsilon\right)\left(1-\epsilon\right)\left(1-\frac{\epsilon}{3}\right)}\right]^n\frac{\Gamma\left(1-\epsilon\right)}{\Gamma\left(1+\frac{\epsilon}{2}\right)\Gamma\left(1-\frac{\epsilon}{2}\right)}\\
\times\,\frac{\Gamma\left(1+\frac{n\epsilon}{2}\right)\Gamma\left(1-\frac{n\epsilon}{2}\right)}{\Gamma\left(1+\frac{n\epsilon}{2}-\frac{\epsilon}{2}\right)\Gamma\left(1-\frac{n\epsilon}{2}-\frac{\epsilon}{2}\right)}\,\frac{\left(1-\frac{n\epsilon}{2}\right)\left(1-\epsilon\right)\left(1-\frac{\epsilon}{3}\right)\left(1-\frac{3\epsilon}{8}-\frac{n\epsilon}{8}\right)}{\left(1-\frac{\epsilon}{2}\right)\left(1-\frac{\epsilon}{2}-\frac{n\epsilon}{2}\right)\left(1-\frac{\epsilon}{4}-\frac{n\epsilon}{4}\right)}\,,
\end{multline}
\begin{multline}\label{poly3}
Y_1(n,\epsilon)=\left(-\frac{4\pi\mu^2}{m_0^2}\right)^{\frac{n\epsilon}{2}}\left[\frac{\Gamma\left(1-\frac{\epsilon}{2}\right)^2\Gamma\left(\frac{\epsilon}{2}+1\right)\left(1-\frac{\epsilon}{2}\right)}{\Gamma\left(1-\epsilon\right)\left(1-\epsilon\right)\left(1-\frac{\epsilon}{3}\right)}\right]^n\\
\times\,\frac{\Gamma\left(1+\frac{n\epsilon}{2}\right)\Gamma\left(1-\frac{n\epsilon}{2}\right)\Gamma\left(1-\epsilon\right)}{\Gamma\left(1-\frac{\epsilon}{2}\right)^3\Gamma\left(1+\frac{\epsilon}{2}\right)}\,\frac{\left(1-\frac{n\epsilon}{2}\right)\left(1-\epsilon\right)\left(1-\frac{\epsilon}{3}\right)\left(1-\frac{\epsilon}{4}\right)}{\left(1-\frac{\epsilon}{2}\right)^2}\,,
\end{multline}
\begin{multline}\label{poly4}
Y_2(n,\epsilon)=\left(-\frac{4\pi\mu^2}{m_0^2}\right)^{\frac{n\epsilon}{2}}\left[\frac{1-\left(\frac{M_0^2}{m_0^2}\right)^{1-\frac{n\epsilon}{2}}}{1-\frac{M_0^2}{m_0^2}}\right]\left[\frac{\Gamma\left(1-\frac{\epsilon}{2}\right)^2\Gamma\left(\frac{\epsilon}{2}+1\right)\left(1-\frac{\epsilon}{2}\right)}{\Gamma\left(1-\epsilon\right)\left(1-\epsilon\right)\left(1-\frac{\epsilon}{3}\right)}\right]^n\times\\
\frac{\Gamma\left(1+\frac{n\epsilon}{2}\right)\Gamma\left(1-\frac{n\epsilon}{2}\right)\Gamma\left(1-\epsilon\right)}{\Gamma\left(1-\frac{\epsilon}{2}\right)^3\Gamma\left(1+\frac{\epsilon}{2}\right)}\,\frac{\left(1-\epsilon\right)\left(1-\frac{\epsilon}{3}\right)}{\left(1-\frac{\epsilon}{2}\right)^2}\,.
\end{multline}

Following the approach of Ref.\cite{PalanquesMestre:1983zy}, we formally expand the functions produced in (\ref{poly1})-(\ref{poly4}) 
as polynomials in $n$ and $\epsilon$\,:
\be\label{ependec}
H_2(n,\epsilon)=\sum_{j=0}^\infty H_2^{(j)}(\epsilon)(n\epsilon)^j\,,
\ee
with analogous series applying to the $H_\phi(n,\epsilon)$, $Y_1(n,\epsilon)$, and $Y_2(n,\epsilon)$ functions.

Inserting \refeq{ependec} into \refeq{loops} and keeping only the poles 
one obtains the following expansion for the fermion self-energy:
\be\label{sigma_exp}
\frac{d\Sigma_2(\slashed{p})_{L,R}}{d\slashed{p}}=\frac{3}{4N_F}\frac{q_{L,R}^2}{q^2}\sum_{n=1}^{\infty}\left(-\frac{2}{3} K_0\right)^n\sum_{j=0}^{n-1} H^{(j)}_2(\epsilon)\frac{n^{j-1}}{\epsilon^{n-j}}\,,
\ee
and analogous expressions apply to $dS(p^2)/dp^2$ and $\Lambda_Y(0,0)$.

We now renormalize the coupling constant, $K_0=Z_3^{-1} K$, by making use of \refeq{gaugeZ3exp} and the binomial series:
\be
K_0^n=Z_3^{-n}K^n=\left(1-\frac{2}{3}\frac{K}{\epsilon}\right)^{-n}K^n=K^n\sum_{i=0}^{\infty}{-n\choose i}\left(-\frac{2}{3}\frac{K}{\epsilon}\right)^i\,.
\ee
Substituting the above into \refeq{sigma_exp} we recast the latter in a useful form:
\be\label{useful}
\frac{d\Sigma_2(\slashed{p})_{L,R}}{d\slashed{p}}=\frac{3}{4N_F}\frac{q^2_{L,R}}{q^2}\sum_{n=1}^{\infty}(-1)^{n}\left(\frac{2K}{3}\right)^n\sum_{j=0}^{n-1} H^{(j)}_2(\epsilon)\frac{n^{j-1}}{\epsilon^{n-j}}\sum_{i=0}^{\infty}(-1)^i{n+i-1\choose i}\left(-\frac{2}{3}\frac{K}{\epsilon}\right)^i\,.
\ee
Finally, after performing a redefinition of the sum parameter, $n\to n-i$, which also indicates that the sum over $i$ is truncated at $i=n-1$,
one gets for \refeq{useful}
\be\label{finalform}
\frac{d\Sigma_2(\slashed{p})_{L,R}}{d\slashed{p}}=\frac{3}{4N_F}\frac{q^2_{L,R}}{q^2}\sum_{n=1}^{\infty}(-1)^{n}\left(\frac{2K}{3}\right)^n\sum_{j=0}^{n-1}\frac{H^{(j)}_2(\epsilon)}{\epsilon^{n-j}}\sum_{i=0}^{n-1}{n-1\choose i}(-1)^i(n-i)^{j-1}\,,
\ee
and analogous expressions can be derived for $dS(p^2)/dp^2$ and $\Lambda_Y(0,0)$.

The crucial observation of Ref.\cite{PalanquesMestre:1983zy} is that the last sum on the right of \refeq{finalform}, 
over the index $i$, is identically zero for all $j>0$. For $j=0$ it yields $-1/n\cdot(-1)^n$. 
Since the only term that gives a non-trivial contribution to \refeq{sigma_exp} is then $H^{(0)}_2(\epsilon)$, we get
\be\label{finalform2}
\frac{d\Sigma_2(\slashed{p})_{L,R}}{d\slashed{p}}=-\frac{3}{4N_F}\frac{q^2_{L,R}}{q^2}\sum_{n=1}^\infty\left(\frac{2K}{3}\right)^n\frac{H_2^{(0)}(\epsilon)}{n\epsilon^{n}}\,,
\ee
with analogous expressions for $dS(p^2)/dp^2$ and $\Lambda_Y(0,0)$. 
The $\epsilon$-dependent functions $H^{(0)}_{2}(\epsilon)$, $H^{(0)}_{\phi}(\epsilon)$, $Y^{(0)}_{1}(\epsilon)$, and 
$Y^{(0)}_{2}(\epsilon)$ can be obtained from Eqs.~(\ref{poly1})-(\ref{poly4}) by setting $n=0$\,:
\bea
H_{2}^{(0)}(\epsilon)&=&\frac{\Gamma\left(1-\epsilon\right)}{\Gamma\left(1+\frac{\epsilon}{2}\right)\Gamma\left(1-\frac{\epsilon}{2}\right)^3}
\frac{\left(1-\epsilon\right)\left(1-\frac{\epsilon}{3}\right)}{\left(1-\frac{\epsilon}{4}\right)}\,,\label{impf1}\\
H_{\phi}^{(0)}(\epsilon)&=&\frac{\Gamma\left(1-\epsilon\right)}{\Gamma\left(1+\frac{\epsilon}{2}\right)\Gamma\left(1-\frac{\epsilon}{2}\right)^3}
\frac{\left(1-\epsilon\right)\left(1-\frac{\epsilon}{3}\right)\left(1-\frac{3\epsilon}{8}\right)}{\left(1-\frac{\epsilon}{2}\right)^2\left(1-\frac{\epsilon}{4}\right)}\,,\label{impf2}\\
Y_{1}^{(0)}(\epsilon)&=&\frac{\Gamma\left(1-\epsilon\right)}{\Gamma\left(1+\frac{\epsilon}{2}\right)\Gamma\left(1-\frac{\epsilon}{2}\right)^3}\frac{\left(1-\epsilon\right)\left(1-\frac{\epsilon}{3}\right)\left(1-\frac{\epsilon}{4}\right)}{\left(1-\frac{\epsilon}{2}\right)^2}\,,\label{impf3}\\
Y_{2}^{(0)}(\epsilon)&=&\frac{\Gamma\left(1-\epsilon\right)}{\Gamma\left(1+\frac{\epsilon}{2}\right)\Gamma\left(1-\frac{\epsilon}{2}\right)^3}\frac{\left(1-\epsilon\right)\left(1-\frac{\epsilon}{3}\right)}{\left(1-\frac{\epsilon}{2}\right)^2}\,.\label{impf4}
\eea

One can expand $H_2^{(0)}(\epsilon)$ in powers of $\epsilon$,
\be\label{overeps}
H_2^{(0)}(\epsilon)=\sum_{i=0}^\infty \widetilde{H}_2^{(i)}\epsilon^i \,,\qquad \widetilde{H}_2^{(0)}=1\,,
\ee
and obtain equivalent formulas for $H^{(0)}_{\phi}(\epsilon)$, $Y^{(0)}_{1}(\epsilon)$, and $Y^{(0)}_{2}(\epsilon)$.
By retaining exclusively the pole terms in $1/\epsilon$ after substituting \refeq{overeps} into \refeq{finalform2}, 
we produce the final form of the renormalized quantities (\ref{loops})-(\ref{loopsL}):
\bea
\frac{d\Sigma_2(\slashed{p})_{L,R}}{d\slashed{p}}&=&-\frac{3}{4N_F}\frac{q^2_{L,R}}{q^2}\sum_{n=1}^\infty\left(\frac{2K}{3}\right)^n\frac{\widetilde{H}_2^{(n-1)}}{n\epsilon}\,,\label{rencos1}\\
\frac{dS(p^2)}{dp^2}&=&\frac{3}{2N_F}\frac{q_S^2}{q^2}\sum_{n=1}^\infty\left(\frac{2K}{3}\right)^n\frac{\widetilde{H}_\phi^{(n-1)}}{n\epsilon}\,,\label{rencos2}\\
\Lambda_Y(0,0) &=&\frac{3\,y}{N_F}\sum_{n=1}^{\infty}\left(\frac{2K}{3}\right)^n\frac{1}{n\epsilon}\left(\frac{q_Lq_R}{q^2}\,\widetilde{Y}_1^{(n-1)}+\frac{1}{4}\frac{q_S^2}{q^2}\,\widetilde{Y}_2^{(n-1)}\right)\label{rencos3}\,.
\eea

We are now ready to calculate the $1/N_F$ gauge contribution to the Yukawa coupling beta function, \refeq{yukNF}.
The coupled Callan-Symanzik equations for the Yukawa coupling of \refeq{bare_yuk} and gauge coupling $K_0=Z_3^{-1}\mu^{\epsilon}K$ read
\bea
0&=&\left(-\frac{1}{Z_{\phi}}\frac{\partial Z_{\phi}}{\partial K}-\frac{1}{Z_{2\,L}}\frac{\partial Z_{2\,L}}{\partial K}-\frac{1}{Z_{2\,R}}\frac{\partial Z_{2\,R}}{\partial K}+\frac{2}{Z_1}\frac{\partial Z_1}{\partial K}\right)\beta_1(K) K+\epsilon+\frac{d\ln\alpha_y}{d\ln\mu}\nonumber\\
0&=&K\epsilon+\beta_1(K) K\left(1-\frac{K}{Z_3}\frac{\partial Z_3}{\partial K}\right)\,,
\eea
which lead to the explicit form of the Yukawa coupling beta function,
\be
\beta_y(K)=K\epsilon\left(\frac{2}{Z_1}\frac{dZ_1}{dK}-\frac{1}{Z_{2\,L}}\frac{dZ_{2\,L}}{dK}-\frac{1}{Z_{2\,R}}\frac{dZ_{2\,R}}{dK}-\frac{1}{Z_\phi}\frac{dZ_\phi}{dK}\right)\,.
\ee

The derivatives of Eqs.~(\ref{rencos1})-(\ref{rencos3}) read
\bea
\epsilon\,\frac{\partial Z_{2\,L,R}}{\partial K}&=&-\frac{1}{2N_F}\frac{q^2_{L,R}}{q^2}\sum_{n=0}^\infty\widetilde{H}_2^{(n)}\cdot\left(\frac{2K}{3}\right)^n=-\frac{1}{2N_F}\frac{q^2_{L,R}}{q^2}\,H_2^{(0)}\left(\frac{2K}{3}\right)\,,\label{render1}\\
\epsilon\,\frac{\partial Z_{\phi}}{\partial K}&=&\frac{1}{N_F}\frac{q_S^2}{q^2}\sum_{n=0}^\infty\widetilde{H}_{\phi}^{(n)}\cdot\left(\frac{2K}{3}\right)^n=\frac{1}{N_F}\frac{q^2_S}{q^2}\,H_{\phi}^{(0)}\left(\frac{2K}{3}\right)\,,\label{render2}\\
\epsilon\,\frac{\partial Z_1}{\partial K}&=&-\frac{2\,y}{N_F}\sum_{n=0}^{\infty}\left(\frac{2K}{3}\right)^n\left(\frac{q_Lq_R}{q^2}\,\widetilde{Y}_1^{(n-1)}+\frac{1}{4}\frac{q_S^2}{q^2}\,\widetilde{Y}_2^{(n-1)}\right)\nonumber\\
 &=&-\frac{2\,y}{N_F}\left[\frac{q_Lq_R}{q^2}\,Y_1^{(0)}\left(\frac{2K}{3}\right)+\frac{1}{4}\frac{q_S^2}{q^2}\,Y_2^{(0)}\left(\frac{2K}{3}\right)\right]\,,\label{render3}
\eea
where the r.h.s. equalities are a direct consequence of \refeq{overeps}.

As the functions $H^{(0)}_{2}(x)$, $H^{(0)}_{\phi}(x)$, $Y^{(0)}_{1}(x)$, and 
$Y^{(0)}_{2}(x)$ are given in Eqs.~(\ref{impf1})-(\ref{impf4}), we can finally write the first order term in the $1/N_F$ expansion of
\refeq{yukNF} in closed form:
\begin{multline}\label{ygrekone}
Y_1(K)K^{-1}=-4\frac{q_L q_R}{q^2}Y_1^{(0)}\left(\frac{2}{3}K\right)-\frac{q_S^2}{q^2}Y_2^{(0)}\left(\frac{2}{3}K\right)\\
+\,\frac{1}{2}\frac{q_L^2+q_R^2}{q^2}H_2^{(0)}\left(\frac{2}{3}K\right)-\frac{q_S^2}{q^2}H_\phi^{(0)}\left(\frac{2}{3}K\right)\,.
\end{multline}

\subsection{Generalization to the non-abelian case\label{sec:nonabel}}

The findings of \refsec{yukabel} can be straightforwardly extended to the case of a non-abelian gauge symmetry 
with interaction strength $g$ ($\alpha_g=g^2/4\pi$), by introducing the appropriate group-theoretical factors. 

We denote by $S_2(R_{\textrm{VL}})$ the Dynkin index of the representation $R_{\textrm{VL}}$ of the 
$N_F$ vector-like fermions, and we redefine the coupling constant as $K=\alpha_gS_2(R_{\textrm{VL}})N_F/\pi$.

In the non-abelian case, then, Eqs.~(\ref{loops})-(\ref{loopsL}) will have to be modified as
\bea
\frac{d\Sigma_2(\slashed{p})_{L,R}}{d\slashed{p}}&=&\frac{3}{4N_F}\frac{C_2(R_{F_{L,R}})}{S_2(R_{\textrm{VL}})}\sum_{n=1}^\infty\left(-\frac{2}{3}K_0\right)^n\frac{1}{n\epsilon ^n}H_2(n,\epsilon)\,,\\
\frac{dS(p^2)}{dp^2}&=&-\frac{3}{2N_F}\frac{C_2(R_S)}{S_2(R_{\textrm{VL}})}\sum_{n=1}^\infty\left(-\frac{2}{3}K_0\right)^n\frac{1}{n\epsilon ^n}H_\phi(n,\epsilon)\,,\\
\Lambda_Y(0,0)&=&-\frac{3y}{N_F}\sum_{n=1}^{\infty}\left(-\frac{2}{3}K_0\right)^n\frac{1}{n\epsilon^n}\left[\frac{C_2(R_{F_L})+C_2(R_{F_R})-C_2(R_S)}{2\,S_2(R_{\textrm{VL}})}Y_1\left(n,\epsilon\right)\right.\nonumber\\
 & &\left.+\,\frac{1}{4}\frac{C_2(R_S)}{S_2(R_{\textrm{VL}})}Y_2\left(n,\epsilon\right)\right]\,,
\eea
where $C_2(R)$ is the quadratic Casimir invariant of a generic representation $R$. Representations $R_S$ and $R_{F_{L,R}}$ 
apply to the scalar $\phi$ and fermions $\psi_{L,R}$ of the Yukawa interaction, \refeq{yukint}. 
The representation of the $N_F$ vector-like fermions, $R_{\textrm{VL}}$, can in principle be different from any of these, but does not have to.

Following step by step the computation outlined in \refsec{yukabel}, 
we derive the following $1/N_F$ gauge contribution to the Yukawa beta function,
\begin{multline}
Y_1(K)K^{-1}=-2\,\frac{C_2(R_{F_L})+C_2(R_{F_R})-C_2(R_S)}{S_2(R_{\textrm{VL}})}Y_1^{(0)}\left(\frac{2}{3}K\right)-\frac{C_2(R_S)}{S_2(R_{\textrm{VL}})}Y_2^{(0)}\left(\frac{2}{3}K\right)\\
+\,\frac{C_2(R_{F_L})+C_2(R_{F_R})}{2\,S_2(R_{\textrm{VL}})}H_2^{(0)}\left(\frac{2}{3}K\right)-\frac{C_2(R_S)}{S_2(R_{\textrm{VL}})}H_\phi^{(0)}\left(\frac{2}{3}K\right)\,.
\end{multline}

By Taylor expanding $Y_1(K)$ around $2K/3=0$, we are now in the position to check that at 1 loop and 2 loops 
the above formula reduces to 
\bea
Y_1^{(\textrm{1 loop})}&=&\frac{K}{S_2(R_{\textrm{VL}})} \left\{-\frac{3}{2}\left[C_2(R_{F_L})+C_2(R_{F_R})\right]\right\},\nonumber\\
Y_1^{(\textrm{2 loop})}&=&\frac{K^2}{S_2(R_{\textrm{VL}})} \left\{\frac{5}{12}\left[C_2(R_{F_L})+C_2(R_{F_R})\right]-\frac{1}{4}C_2(R_S)\right\},
\eea
which lead, after explicitly expressing $K$ in terms of $g$, to the well known forms of Ref.\cite{Machacek:1983fi}.

\section{Properties and applications\label{sec:properties}}

\begin{figure}[t]
\centering
\includegraphics[width=0.6\textwidth]{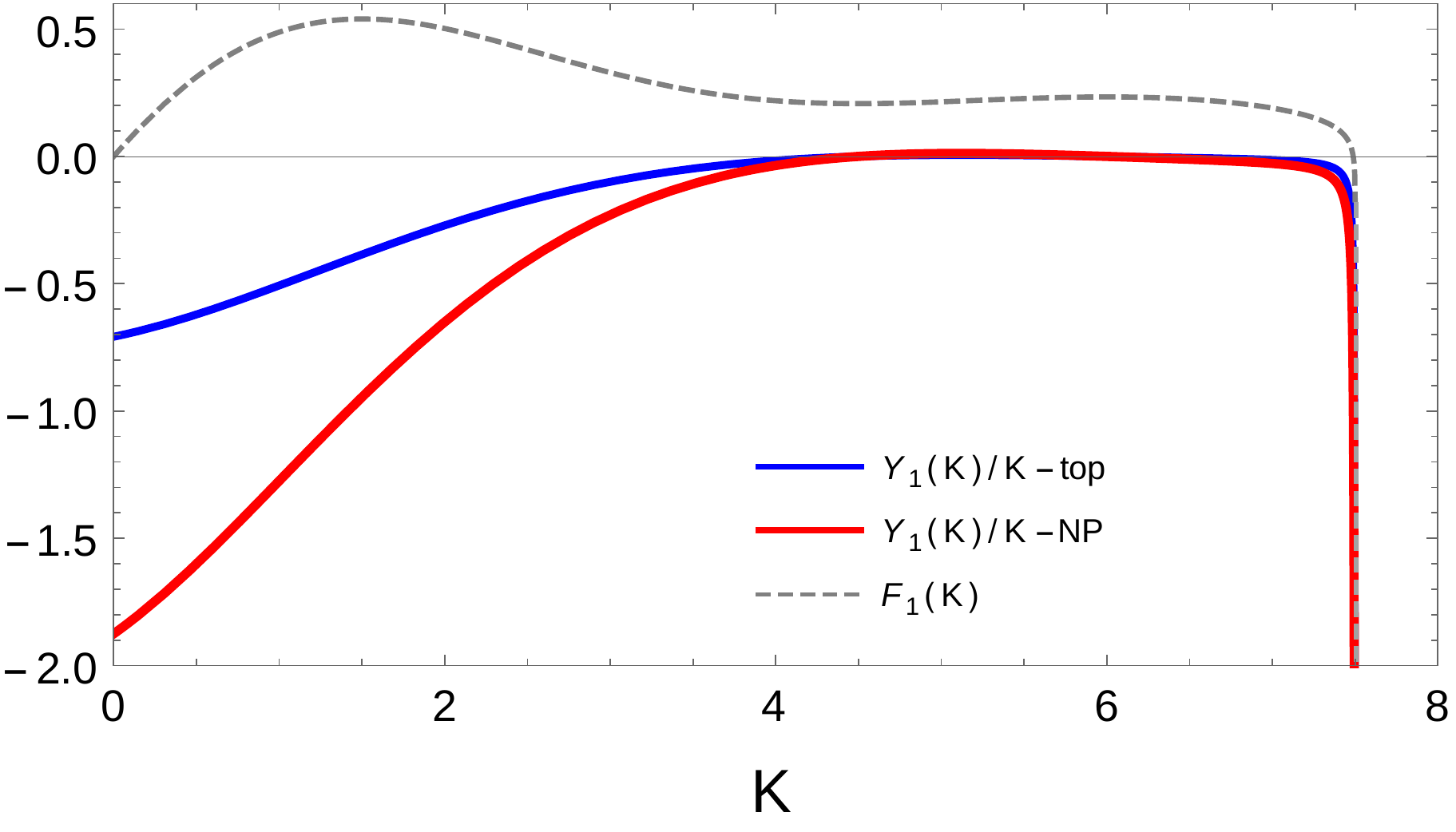}
\caption{A plot of the function $Y_1(K)/K$ defined in \refeq{ygrekone} for the case of the top quark Yukawa coupling (solid blue) and 
a model of new physics defined in the text (solid red). The dashed gray line shows the function $F_1(K)$, given in \refeq{ef1}, 
for comparison.}
\label{fig:singular}
\end{figure}

To illustrate the properties of the resummed gauge contributions to the gauge and Yukawa beta functions, we plot 
in \reffig{fig:singular} the dependence of $Y_1(K)/K$ on the value of the rescaled gauge coupling $K$. 
We limit ourselves here to the abelian case, \refeq{ygrekone}. In other words, we assume that 
the new $N_F$ vector-like fermions responsible for the tower of resummed bubble diagrams 
are charged under the U(1) gauge group only, with $q=1$.

We show with a solid blue line the case of the SM top quark Yukawa, 
and with a solid red line an example case for a model of 
new physics (indicated with NP), 
in which we assume there exist a new Yukawa coupling between the SM muon (with U(1) charge $q_L=-1/2$), 
a new inert SU(2) scalar doublet with $q_S=1/2$, and an SU(2)-singlet fermion with U(1) charge $q_R=-1$,
all of them color-neutral. 
This model has been chosen for illustrative purposes, 
but it was shown in several papers, e.g.,\cite{Belanger:2015nma,Arnan:2016cpy,Kowalska:2017iqv}, that such a construction 
can be used  to provide a good fit, through loop contributions, 
to the muon $g-2$ measurement and, with addition of an appropriate colored sector,
to the LHCb $b\rightarrow s$ anomalies, for reasonable choices of the mass of the new particles, 
as long as the new Yukawa coupling is substantial, $y_{\textrm{NP}}\gsim 1-2$.  

The dashed gray line shows the behavior of the function $F_1(K)$, defined in \refeq{ef1}. 
It was first observed in\cite{PalanquesMestre:1983zy} that $F_1(K)$ exhibits a singularity for $K=15/2$, 
which then determines the radius of convergence for the $\beta_1(K)$ expansion. 
As anticipated, $Y_1(K)/K$ presents a pole at the same value of $K$. 

A straightforward consequence of the pole structure of $F_1(K)$ is the possibility of generating 
a non-trivial UV fixed point in the RG flow of the gauge coupling\cite{Holdom:2010qs}. 
Indeed, in the vicinity of the singularity at $K=15/2$, $F_1(K)$ becomes negative and the beta function  $\beta_1$ eventually vanishes. 
As a result, the gauge coupling asymptotically approaches the value
\be\label{fp_gauge}
\alpha_1^{\ast}=\frac{15}{2}\frac{\pi}{N_F\,q^2}\,.
\ee
This idea was brought forward in\cite{Mann:2017wzh} 
as a mechanism for avoiding the Landau pole in the running of the hypercharge gauge coupling of the SM. 
Interestingly, we find that $Y_1(K)/K$ is also negative as $K$ approaches the pole, which has consequences of crucial importance
for the asymptotic UV behavior of the theory.

\begin{figure}[t]
\centering
\includegraphics[width=0.6\textwidth]{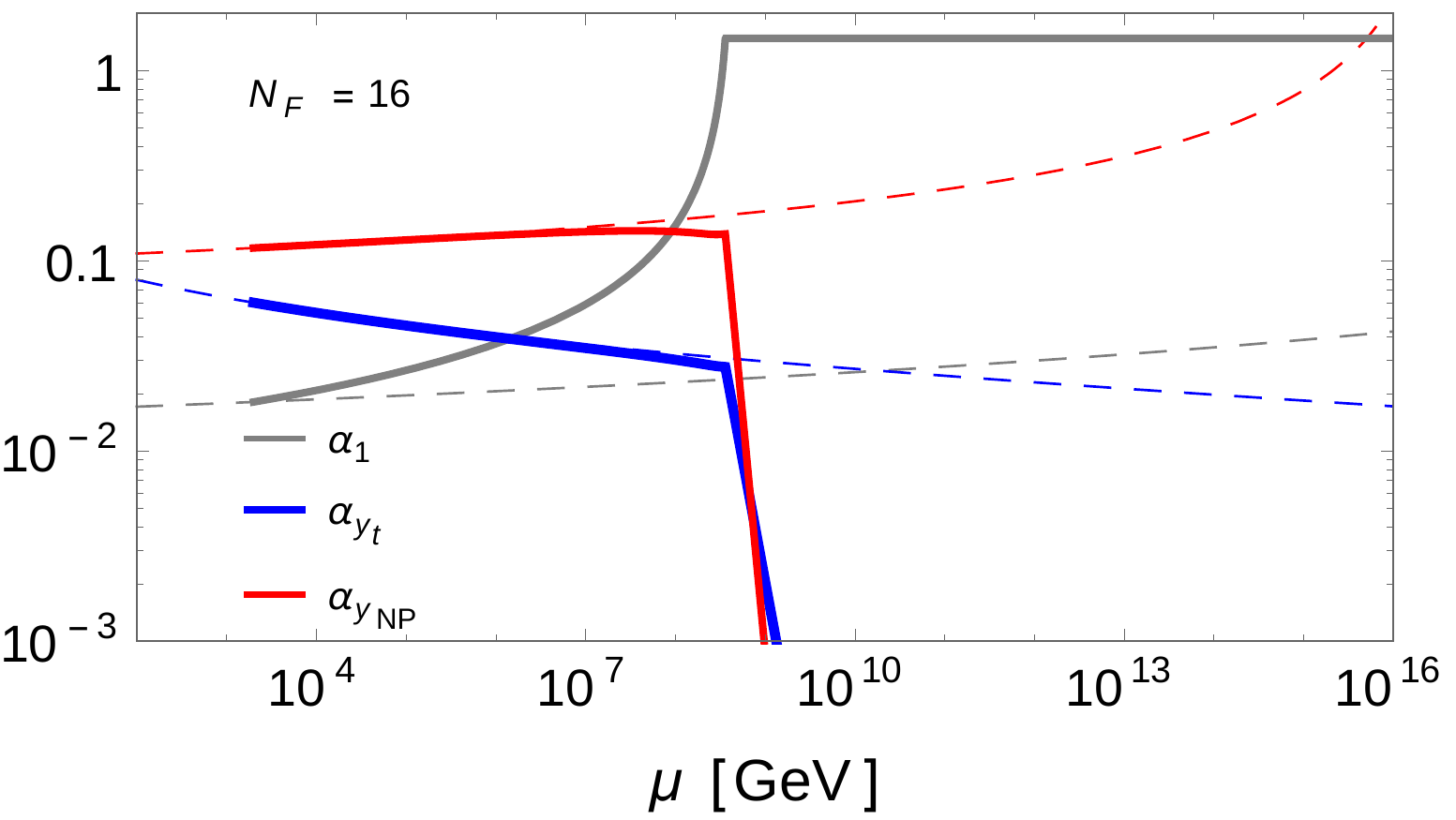}
\caption{Modified running of the hypercharge U(1) coupling $\alpha_1$ (gray solid), top Yukawa coupling $\alpha_{y_t}$ (blue solid), 
and the Yukawa coupling $\alpha_{y_{\textrm{NP}}}$ of a model of new physics defined in the text (red solid) in the presence of $N_F=16$ vector-like fermions charged under U(1).
The common mass of the vector-like fermions is set at 2 TeV.
For each coupling, the corresponding 1-loop running without the large $N_F$ enhancement 
is shown with a dashed line of the same color. }
\label{fig:run}
\end{figure}

We illustrate the behavior of the beta functions in \reffig{fig:run}. 
Solid lines of different colors show the running of the different beta functions in the presence of $N_F=16$ vector-like fermions with charge $q=1$ and a common mass $m_{\textrm{VL}}\approx2\tev$. 

In gray, we present the running of $\alpha_1$. The $N_F$ vector-like fermions modify the 
SM one-loop beta function (dashed line) by introducing extra contributions,
\be\label{rges_gau}
\frac{d\alpha_1}{d\ln\mu}=\frac{\alpha_1^2}{4\pi}\left[\frac{41}{3} + \frac{8}{3}N_F + \frac{8}{3}F_1\left(\frac{\alpha_1 N_F}{\pi}\right)\right]\,,
\ee
which push the coupling to the UV fixed point appearing in the solid line. 

Under the same assumptions, Yukawa couplings are also modified. For instance, for the top quark we get
\be
\frac{d\alpha_{y_t}}{d\ln\mu}=\frac{\alpha_{y_t}}{4\pi}\left[9\alpha_{y_t}-16\alpha_3-\frac{9}{2}\alpha_2+ \frac{4\pi}{N_F}Y_1\left(\frac{\alpha_1 N_F}{\pi}\right)\right]\,,
\ee
which changes the running of the coupling from its familiar SM path, shown in dashed blue in \reffig{fig:run}, to the solid blue line.  
One can observe that since the $Y_1$ function introduces a large negative contribution, the renormalized value of the Yukawa coupling 
begins to decrease much more rapidly when the gauge coupling approaches its fixed point. 

This fact can have interesting consequences for those new physics models in which large Yukawa 
couplings are desirable to better accommodate experimental anomalies. In that case, 
the presence of high-scale vector-like fermions can make the gauge and Yukawa sectors of the theory asymptotically safe. 
For instance, one can consider again the model introduced above, in \reffig{fig:singular}, featuring a new physics Yukawa coupling 
between the SM muon and two new fields, $\mathcal{L}\supset y_{\textrm{NP}}\,\bar{\psi}_{\mu,L}\,\phi\,\psi_R$\,.
We choose in \reffig{fig:run} an initial value of $y_{\textrm{NP}}$ that leads to the Landau pole being below the Planck scale 
(dashed red line) if only 1-loop contributions to the beta function were taken into account. 
However, in the large-$N_F$ limit, the Yukawa coupling beta function is modified as
\be
\frac{d\alpha_{y_{\textrm{NP}}}}{d\ln\mu}=\frac{\alpha_{y_{\textrm{NP}}}}{4\pi}\left[5\,\alpha_{y_{\textrm{NP}}}-\frac{9}{2}\alpha_2+ \frac{4\pi}{N_F}Y_1\left(\frac{\alpha_1 N_F}{\pi}\right)\right]\,,
\ee
and the model does not suffer from a Landau pole any longer (solid red line).

Before we conclude, we would like to recall an important point of difference between the abelian case discussed above and its 
non-abelian counterpart, introduced in \refsec{sec:nonabel}. As was pointed out in\cite{Holdom:2010qs}, following the calculation of\cite{Gracey:1996he}, the non-abelian gauge coupling beta function, $\beta_g$, features a logarithmic singularity at $K=3$ (where $K=\alpha_gS_2(R_{\textrm{VL}})N_F/\pi$). Thus, in the non-abelian case the gauge beta function reaches its fixed point 
well before the function $Y_1(K)$ has the chance of approaching its drastic drop at $K\rightarrow 15/2$, 
see \reffig{fig:singular}. As a consequence, the running of the Yukawa coupling beta function in the non-abelian case continues along a 
path similar to the one defined by the standard 1-loop contributions, and it is not much affected by the resummation.
Note that the discrepancy between the position of the pole in the non-abelian gauge beta function and the one in the anomalous dimension 
of the fermion mass was also discussed in Ref.\cite{Antipin:2017ebo}.   

\section{Conclusions\label{sec:summary}}

In this paper we calculated the gauge contributions to the $1/N_F$ expansion of the beta function of a generic Yukawa coupling. 
To this end, we summed an infinite series of diagrams with ever increasing number of vacuum-polarization bubbles to obtain an 
analytical closed form. The resummed contribution exhibits a simple pole, which in the abelian case sits at the same value 
of the gauge coupling for which a logarithmic singularity in the gauge beta function has been long known to exist. 
As a direct consequence, close to the singularity the RG evolution of the Yukawa coupling is strongly affected. 
In the abelian case, in particular, it can become asymptotically free even for low-energy boundary conditions that would lead to a 
Landau pole close to the TeV-scale if only the standard finite-loop contributions were taken into account. 

This feature could have important phenomenological consequences. Many new physics scenarios require non-perturbative Yukawa couplings to maximally enhance loop contributions to various low-energy measurements, like the muon $g-2$ discrepancy or LHCb flavor anomalies. 
The resummed impact of $N_F$ vector-like fermions on the evolution of both the gauge and Yukawa couplings offers a way to construct 
a UV-completion in which these two sectors of the theory remain asymptotically safe. 
In the purely non-abelian case, however, this mechanism is not effective, due to a mismatch 
between the position of the pole for the gauge and Yukawa coupling beta functions, which mirrors the equivalent mismatch 
of the pole position in the fermion anomalous mass dimension with respect to the gauge beta function.  

The necessary next step to prove the full asymptotic safety of the gauge-Yukawa theory should be 
to consider the impact of the same infinite series of vacuum-polarization diagrams on the scalar quartic coupling. Incidentally,
this might also be an issue of high importance for the stability of the scalar potential. 
Additionally, the calculation could be extended to accommodate other assumptions, one of which could be that the scalar field coupled uniformly to all $N_F$ vector-like fermions. 
In such a case, one expects an $N_F$ enhancement of the scalar propagator, 
in analogy to what was discussed here for the gauge bosons, and additional combinatorial factors. 
These ramifications notwithstanding, we find it encouraging that already at the level of the pure gauge coupling contributions presented here, the resummation can possibly 
lead to novel phenomenological applications.

\bigskip
\noindent \textbf{Acknowledgments}
\medskip

\noindent We would like to thank Luc Darm\'e for his comments on the manuscript. 
EMS would like to thank the Physics Department at TU Dormund for hospitality and the Alexander von Humboldt Foundation 
for support during the initial stages of this work.

\appendix
\section{Vacuum polarization}\label{appen:vac}

We recall in this appendix a few notions on the gauge dependence of the photon propagator and briefly sketch the steps leading 
to the photon self-energy 1-particle irreducible (1PI) 
correction presented in \refeq{pizen}, as the integrals of Eqs.~(\ref{fermion})-(\ref{vertex}) are calculated following similar lines.

The bare photon propagator, $\Delta_{B\,\mu\nu}$ in a general Lorentz invariant gauge reads,
\be
\Delta_{B\,\mu\nu}=\frac{g_{\mu\nu}-\xi(k^2)k_{\mu}k_{\nu}/k^2}{k^2-i\epsilon}\,,
\ee
in terms of a gauge-fixing parameter $\xi(k^2)$. On the other hand, the complete photon propagator, 
obtained after resummation of 1PI graphs,
\be
\Delta_{\mu\nu}'=\Delta_{B\,\mu\nu}+\Delta_{B\,\mu\rho}\Pi^{\rho\sigma}(k)\Delta_{B\,\sigma\nu}+...\,,
\ee 
where the sum of all 1PI insertions
is parameterized as
\be
i\Pi^{\rho\sigma}(k)=i\left(g^{\rho\sigma}k^2-k^{\rho}k^{\sigma}\right)\Pi(k^2)\,,
\ee
is given by (see, e.g., Chapter~10 of\cite{Weinberg:1995mt})
\be\label{fullWein}
\Delta_{\mu\nu}'=\frac{g_{\mu\nu}-\tilde{\xi}(k^2)k_{\mu}k_{\nu}/k^2}{\left(k^2-i\epsilon\right)\left[1-\Pi(k^2)\right]}\,.
\ee

Equation~(\ref{fullWein}) is expressed in terms of the \textit{renormalized} gauge fixing term,
\be
\tilde{\xi}(k^2)=\xi(k^2)\left[1-\Pi(k^2)\right]+\Pi(k^2)\,,
\ee 
where, as we shall see shortly, $\Pi(k^2)$ takes at one loop the explicit form presented in \refeq{pizen}. 

In the Feynman gauge, which we adopt in this work, $\xi(k^2)=0$, so that $\tilde{\xi}(k^2)=\Pi(k^2)$.  
Equation~(\ref{fullWein}) can then be expanded back in a power series of $\Pi(k^2)$ to yield 
\be\label{expanWein}
\Delta_{\mu\nu}'=\frac{g_{\mu\nu}}{k^2}\sum_{n=0}^{\infty}\Pi^n(k^2)-\frac{k_{\mu}k_{\nu}}{k^4}\sum_{n=0}^{\infty}\Pi^{n+1}(k^2)\,.
\ee
Note that at the $n$-th order in perturbation theory, in dimensional regularization $\Pi^n(k^2)$ assigns to the loop integral a pole contribution of the order of $1/\epsilon^n$,
whereas $\Pi^{n+1}(k^2)$ produces a higher order term, in $1/\epsilon^{n+1}$, which therefore does not contribute to the beta function, see Eqs.~(\ref{rencos1})-(\ref{rencos3}).
Thus, we can safely neglect the second term of \refeq{expanWein} when calculating the integrals  of Eqs.~(\ref{fermion})-(\ref{vertex}). 

We conclude this appendix by explicitly calculating \refeq{pizen}. In $d=4-\epsilon$ dimensions, we write at one loop
\be\label{1PIstart}
i\Pi^{\rho\sigma}(k)=-(-ig_{1,0}\,q)^2 N_F \mu^{\epsilon}\int\frac{d^d l}{(2\pi)^d}\textrm{Tr}\left[\gamma^{\rho}\frac{i(\slashed{l}+m_0)}{l^2-m_0^2}\gamma^{\sigma}\frac{i(\slashed{l}-\slashed{k}+m_0)}{(l-k)^2-m_0^2}\right]\,.
\ee

After reducing the numerator in \refeq{1PIstart} by making use of Dirac matrix identities, and simplifying the denominator with
the usual Feynman parameter $x$, 
\be
\frac{1}{A^{\alpha}B^{\beta}}=\frac{\Gamma\left(\alpha+\beta\right)}{\Gamma\left(\alpha\right)\Gamma\left(\beta\right)}\int_0^1dx\,\frac{x^{\alpha-1}\left(1-x\right)^{\beta-1}}{\left[Ax+B\left(1-x\right)\right]^{\alpha+\beta}}\,,
\ee
one gets
\be\label{1PImed}
i\Pi^{\rho\sigma}(k)=4(-ig_{1,0}\,q)^2 N_F \mu^{\epsilon}\int_0^1 dx\int\frac{d^d l}{(2\pi)^d}\frac{l^{\rho}(l^{\sigma}-k^{\sigma})+l^{\sigma}(l^{\rho}-k^{\rho})-g^{\rho\sigma}\left[l\cdot(l-k)-m_0^2\right]}{\left[l^2-2l\cdot k\,x+k^2x-m_0^2\right]^2}\,.
\ee

By redefining $\tilde{l}=l-x\,k$ and $\Delta=m_0^2-x(1-x)k^2$, and neglecting terms linear in $\tilde{l}$, one modifies \refeq{1PImed} into
\begin{multline}\label{1PIfin}
i\Pi^{\rho\sigma}(k)=4(-ig_{1,0}\,q)^2 N_F \mu^{\epsilon}\int_0^1 dx\int\frac{d^d \tilde{l}}{(2\pi)^d}\,\left[\frac{2\,\tilde{l}^{\rho}\tilde{l}^{\sigma}-g^{\rho\sigma}\tilde{l}^2}{\left(\tilde{l}^2-\Delta\right)^2}\right.\\
\left.+\,\frac{x\left(x-1\right)\left(2\,k^{\rho}k^{\sigma}-g^{\rho\sigma}k^2\right)+g^{\rho\sigma}m_0^2}{\left(\tilde{l}^2-\Delta\right)^2}\right]\,.
\end{multline}

Finally, after recalling that $\tilde{l}^{\rho}\tilde{l}^{\sigma}=\tilde{l}^2\,g^{\rho\sigma}/d$, the two pieces of 
\refeq{1PIfin} can be Wick-rotated and easily integrated to give
\be
4i(-ig_{1,0}\,q)^2 N_F\mu^{\epsilon}\frac{-g^{\rho\sigma}}{(4\pi)^{2-\frac{\epsilon}{2}}}\,\Gamma\left(\frac{\epsilon}{2}\right)\int_0^1dx \left(\frac{1}{\Delta}\right)^{-1+\frac{\epsilon}{2}}
\ee
for the first, and 
\be
4i(-ig_{1,0}\,q)^2 N_F\mu^{\epsilon}\frac{\left(2k^{\rho}k^{\sigma}-g^{\rho\sigma}k^2\right)}{(4\pi)^{2-\frac{\epsilon}{2}}}\,\Gamma\left(\frac{\epsilon}{2}\right)\int_0^1dx\,x\left(x-1\right) \left(\frac{1}{\Delta}\right)^{\frac{\epsilon}{2}}
\ee
for the second, where in the latter we only considered the limit $m_0^2\ll k^2$. 

It is now straightforward to perform the integrations in $x$, which yield, again in the limit $m_0^2\ll k^2$,
\bea
\int_0^1dx \left(\frac{1}{\Delta}\right)^{-1+\frac{\epsilon}{2}}&=&-k^2\left(-k^2\right)^{-\frac{\epsilon}{2}}\frac{\Gamma\left(2-\frac{\epsilon}{2}\right)^2}{\Gamma\left(4-\epsilon\right)}\\
\int_0^1dx\,x\left(x-1\right) \left(\frac{1}{\Delta}\right)^{\frac{\epsilon}{2}}&=&-\left(-k^2\right)^{-\frac{\epsilon}{2}}\frac{\Gamma\left(2-\frac{\epsilon}{2}\right)^2}{\Gamma\left(4-\epsilon\right)}\,,
\eea
and lead to the final form for $i\Pi^{\rho\sigma}(k)$:
\be\label{1PIfinal}
i\Pi^{\rho\sigma}(k)=i\left(g^{\rho\sigma}k^2-k^{\rho}k^{\sigma}\right)(-ig_{1,0}\,q)^2 
N_F\frac{8}{\left(4\pi\right)^2}\left(-\frac{4\pi\mu^2}{k^2}\right)^{\frac{\epsilon}{2}}\frac{\Gamma\left(\frac{\epsilon}{2}\right)\Gamma\left(2-\frac{\epsilon}{2}\right)^2}{\Gamma\left(4-\epsilon\right)}\,.
\ee

After a redefinition of $K_0=\alpha_{1,0}\,q^2 N_F/\pi$, \refeq{1PIfinal} is easily recast into \refeq{pizen}.

\bibliographystyle{JHEP}
\bibliography{KE4}
\end{document}